\documentclass[11pt,a4paper]{article}
\pdfoutput=1

\usepackage{jheplike}
\usepackage{amsmath,amsfonts,amssymb}
\usepackage{array}

\hypersetup{pdftitle={Instanton induced Yukawa couplings from distant E3 and E(-1) instantons}, 
pdfauthor={Mark D. Goodsell, Lukas T. Witkowski}, pdfsubject={String theory model building},
bookmarksopen, bookmarksnumbered, bookmarksopenlevel=2}

\newcommand{\be}{\begin{equation}}
\newcommand{\ee}{\end{equation}}

\newcommand{\tha}[3]{\vartheta[\!\begin{array}{c}{\phantom{}\vspace{-1.2mm}\scriptstyle#1}%
                        \\[-2.0mm]{\scriptstyle #2}\end{array}\!]( { #3} )}

\def\nn{\nonumber}
\def\pd{\partial}
\def\d{\textrm{d}}
\def\ap{\alpha^{\prime}}
\def\Re{\textrm{Re}}
\def\zb{\bar{z}}

\def\bra{\langle}
\def\ket{\rangle}
\def\ov{\overline}

\title{Instanton induced Yukawa couplings from distant E3 and E(-1) instantons}

\author[1, 2, a]{Mark D.~Goodsell\note[a]
{\href{mailto:mark.goodsell@lpthe.jussieu.fr}
{mark.goodsell@lpthe.jussieu.fr}}}
\author[3, b]{and Lukas T.~Witkowski\note[b]
{\href{mailto:L.Witkowski@ThPhys.Uni-Heidelberg.de}
{L.Witkowski@ThPhys.Uni-Heidelberg.de}}}

\affiliation[1]{Sorbonne Universit\'es, UPMC Univ Paris 06, UMR 7589, LPTHE, F-75005, Paris, France \vspace{0.1cm}}
\affiliation[2]{CNRS, UMR 7589, LPTHE, F-75005, Paris, France \vspace{0.1cm}}
\affiliation[3]{Institute for Theoretical Physics, University of Heidelberg, 
Philosophenweg 19, 69120 Heidelberg, Germany \vspace{0.1cm}}

\abstract{
We calculate non-perturbative contributions to Yukawa couplings on D3-branes at orbifold singularities due to E3 and fractional E(-1) instantons which do not intersect the visible sector branes.  While distant E3 instantons on bulk cycles typically contribute to Yukawa couplings, we find that distant fractional E(-1) can also give rise to new Yukawa couplings. However, fractional E(-1) instantons only induce Yukawa couplings if they are located at a singularity which shares a collapsed homologous two-cycle with the singularity supporting the visible sector. The non-perturbative contributions to Yukawa couplings exhibit a different flavour structure than the tree-level Yukawa couplings and, as a result, they can be sources of flavour violation. This is particularly relevant for schemes of moduli stabilisation which rely on superpotential contributions from E3 instantons, such as KKLT or the Large Volume Scenario. As a byproduct of our analysis, we shed some new light on the properties of annulus diagrams with matter field insertions in stringy instanton calculus.  
}

\begin{document}

\maketitle

\section{Introduction}
\label{Introduction}
A study of the phenomenology of compactifications of string theory is not complete without addressing non-perturbative effects. In the context of type II string theory Euclidean D-brane instantons or gaugino condensation on D-branes can induce non-perturbative contributions to the superpotential \cite{9604030} with important consequences for moduli stabilisation and particle physics phenomenology. In particular, non-perturbative superpotential contributions are a crucial ingredient of successful schemes of K\"ahler moduli stabilisation in type IIB string theory \cite{KKLT, LVS}. Furthermore, Euclidean D-brane instantons can give rise to ``new'' visible sector couplings which are otherwise forbidden perturbatively \cite{0609191, 0609213, 0610003, Argurio:2007vqa, 07040784} (see also \cite{0609211}).\footnote{Euclidean D-branes also give rise to a string theory realisation of gauge instantons and there is an extensive literature on this subject. We refer readers to the review \cite{09023251} and references therein. }

This last point is of interest to phenomenologists for various reasons. Non-perturbatively generated couplings of visible sector fields $\Phi_i$ are suppressed by the instanton action
\be
W \supset \prod_{i=1}^{n} \Phi_i \ e^{-S_{inst}}
\ee
allowing for an implementation of hierarchical Yukawa couplings \cite{0612110, 07071871, 07111316, 08111583, 09053379} or the seesaw mechanism for neutrino masses \cite{0609191, 0609213, 0703028, 07041079}. ``Power-towers'' of instantons \cite{Blumenhagen:2008ji,Blumenhagen:2012kz} may be relevant for cosmology \cite{Cicoli:2011ct,Cicoli:2012tz}. Further, it was suggested that stringy instantons may allow for de Sitter vacua in type IIB compactifications \cite{12031750}.

It is well known that instantons can induce perturbatively forbidden couplings if the instanton intersects visible sector branes. However, visible sector couplings can also be generated if the instanton and the visible sector are spatially separated. In type IIA string theory it was shown that distant E2 instantons give rise to Yukawa couplings on D6-branes, thus allowing for a solution to the ``rank one'' problem \cite{0612110}. In type IIB string theory distant E3 instantons can induce new visible sector operators on D3 and D7-branes \cite{08062291, 09105496}. Corrections to Yukawa couplings from E3 instantons also allow for a realistic Yukawa hierarchy in F-theory models \cite{12116529, 13078089, 150302683}.

There is also a geometric understanding for the generation of visible sector operators due to distant E3 instantons \cite{0607050}. To be specific, let us consider a visible sector on the worldvolume of a D3-brane. If the E3 instanton exhibits the correct number of fermionic zero modes the D3 theory will exhibit a superpotential term whose size is determined by the volume of the 4-cycle $\Sigma$ wrapped by the E3:
\be
|W_{D3}^{np}| \sim e^{-2 \pi \textrm{Vol}_{\Sigma}} \ .
\ee
The generation of matter operators in the D3-theory can then be understood as follows. The presence of the D3-brane backreacts on the geometry and modifies the volume of $\Sigma$. When the D3-brane moves w.r.t.~the E3 location, the strength of this backreaction changes and so does $\textrm{Vol}_{\Sigma}$. As a result $W_{D3}^{np}$ depends on the D3-position $z_{D3}$. After identifying fluctuations in the D3-positions with D3 matter fields one concludes that distant E3 instantons can indeed give rise to new D3 matter interactions.

While the geometric picture gives a compelling conceptual understanding for the generation of D3 interactions due to E3 instantons, explicit calculations are much harder. Using a geometric ansatz it was shown how D3-branes modify the volume of a bulk 4-cycle in a toroidal orbifold \cite{0607050}, confirming an earlier result obtained by worldsheet CFT methods \cite{0404087}. Further investigations considered D3-branes at singularities or in warped throat geometries, with E3 instantons wrapping non-compact bulk cycles \cite{0607050, 09105496}. 

However, physically interesting situations also involve E3 instantons wrapped on blow-up cycles, which are a crucial ingredient for moduli stabilisation along the lines of the Large Volume Scenario \cite{LVS}. The question whether such E3 configurations give rise to visible sector operators on distant D3-branes is not clear. Answering this question using the geometric ansatz mentioned above is difficult due to the poor understanding of metrics on Calabi-Yau spaces. 

Worldsheet CFT techniques offer an alternative way of studying these effects. In \cite{12071103} worldsheet CFT methods were used to study non-perturbative contributions to Yukawa couplings from gaugino-condensation on D7-branes in a toroidal orbifold setup. In this paper we will perform a similar analysis for distant instantons. While our analysis will be limited to toroidal orbifold backgrounds, we can go beyond non-perturbative effects from E3-branes on bulk cycles. In particular, we will also study new Yukawa couplings induced by distant fractional E(-1) instantons. While fractional E(-1) instantons are interesting in themselves, they can also help shed light on E3 instantons on blow-up cycles. In the geometric regime fractional E(-1) instantons can be understood as E3/E1/E(-1) bound states. In turn, E(-1) instantons can also be seen as a toy model of E3 instantons on a blow-up cycle (in the limit where the cycle collapses to zero size). 

In this paper we will hence calculate non-perturbative contributions to Yukawa couplings from E3-branes on bulk cycles as well as from fractional E(-1) instantons. The first calculation will reproduce known results \cite{08062291, 09105496} and will serve as a check of our methods. The calculation for distant fractional E(-1) instantons has not been performed so far. We will also comment on how we expect our results for fractional E(-1) instantons to generalise to E3 instantons in smooth geometries.

\subsubsection*{Superpotential desequestering}

The existence of non-perturbative contributions to Yukawa couplings will have important implications for the phenomenology of supersymmetry breaking. If present, such new Yukawa couplings take the form
\be
\label{eq:npYuk}
W \supset Y_{\alpha \beta \gamma}^{np} C^{\alpha} C^{\beta} C^{\gamma} = y_{\alpha \beta \gamma}^{np} C^{\alpha} C^{\beta} C^{\gamma} e^{- a T} \ ,
\ee
and lead to direct cross couplings between visible sector fields $C^{\alpha}$ and the sector of K\"ahler moduli $T$.
In the most popular schemes of moduli stabilisation in type IIB supersymmetry is broken dominantly by the K\"ahler moduli which acquire the largest F-terms. Using the standard supergravity formulae \cite{9707209} it follows that the above Yukawa couplings \eqref{eq:npYuk} contribute to soft A-terms (before canonical normalisation) as 
\be 
\label{eq:newA}
\delta A_{\alpha \beta \gamma}^{\prime} \sim e^{K/2} F^T \partial_T Y_{\alpha \beta \gamma}^{np} \sim - e^{K/2} F^T y_{\alpha \beta \gamma}^{np} a e^{-a T} \ .
\ee
While non-perturbative contributions to A-terms would typically give rise to subleading effects, they can become important in setups with \emph{sequestered} supersymmetry breaking. Sequestering refers to the suppression of soft supersymmetry breaking terms below the gravitino mass, and it is an attractive property for semi-realistic models in type IIB (see e.g.~\cite{0703105, 10121858} for discussions of sequestering in string theory). For example, sequestering admits a hierarchy between moduli masses and masses of superpartners, thus allowing for TeV scale SUSY while keeping moduli heavy enough to avoid the Cosmological Moduli Problem.\footnote{See \cite{Coughlan:1983ci} for the pioneering work on the Cosmological Moduli Problem in supergravity and \cite{9308292, 9308325} for early work on this topic in string theory.} The relevance of new contributions to Yukawa couplings \eqref{eq:npYuk} becomes evident if one of the fields $C^{\alpha}$ is a Higgs field. In this case the A-terms \eqref{eq:newA} generate new contributions to soft masses for scalars after electroweak symmetry breaking. For the case of sequestered supersymmetry breaking, where all other contributions to soft masses are suppressed, any new contributions due to non-perturbative effects are potentially important and cannot be ignored. In this sense the existence of non-perturbative contributions to Yukawa couplings is an obstacle to sequestering and has been termed ``superpotential desequestering'' in \cite{10121858}.

In addition, non-perturbative contributions to A-terms are a potential source of flavour violation and their presence can lead to strong constraints on the string model. It is well known that soft A-terms can lead to flavour violating interactions if these terms are not proportional to the Yukawa couplings. The important observation now is, that there is no reason \emph{a priori} why $y_{\alpha \beta \gamma}^{np}$ (and hence $\delta A_{\alpha \beta \gamma}^{\prime}$) should exhibit the same flavour structure as the tree-level Yukawa couplings, making \eqref{eq:newA} a possible source of flavour violation. The resulting constraints are particularly stringent for models with sequestered supersymmetry breaking. For models based on the Large Volume Scenario (LVS) the absence of excessive flavour violation leads to an upper bound for the compactification volume, which for sequestered realisations of the LVS (e.g.~\cite{09063297, 14091931}) excludes phenomenologically interesting parameter space \cite{10121858}. 

\subsubsection*{Reading guide}
The paper is structured as follows. In section \ref{sec:setup} we set up the calculation by reviewing the instanton calculus and outline our expectations based on supersymmetry. The calculation is performed in section \ref{sec:calculation} in the context of IIB theories with branes at singularities. The result is discussed in section \ref{sec:discussion}, where we show how the properties due to supersymmetry generalise to generic backgrounds. We also comment on the flavour structure (sec.~\ref{sec:flavour}). In sec.~\ref{sec:conclusions} we summarise our results.
 
\section{Non-perturbative superpotential contributions in type II string theory}
\label{sec:setup}

\subsection{Effective supergravity theory}

We are interested in extracting holomorphic couplings in the effective supergravity theory; these therefore benefit from the powerful supersymmetric non-renormalisation theorems. These state that the Wilsonian superpotential receives no perturbative corrections, and that the gauge kinetic function receives perturbative corrections only at one loop. However, the symmetries of the theory can also be used to strongly constrain the form of the \emph{non-perturbative} corrections; this was stated succinctly in the IIA context in \cite{Akerblom:2007uc}, which we shall now recall. 

\subsubsection*{IIA}

In type IIA orientifolds with D6 branes, there are $h^{2,1}+1$ chiral superfields which define the complex structure moduli and the dilaton, together written as 
\begin{align}
U_I \equiv \frac{1}{\ell_s^3} \bigg[ \int_{A_I} \mathrm{Re}(S\hat{\Omega}_3) - i \int_{A_I} C_3 \bigg]
\end{align}
where $A_I$ is a basis of three-cycles, $I=0,1,...,h^{2,1}$, $\hat{\Omega}_3$ is the normalised holomorphic three-form and $C_3$ is the R-R three-form. $S$ is the four-dimensional dilaton, rescaled from its ten-dimensional version by the square root of the volume of the Calabi-Yau compactification manifold (note that the definition for $I=0$ is essentially tautological, so the dilaton is included among the complex-structure moduli). Hence the real parts of the $U_I$ are proportional to the string coupling.
Next there are the K\"ahler moduli:
\begin{align}
T_i \equiv \frac{1}{\ell_s^2} \int_{C_i} J_2 - i B_2
\end{align}
where the $C_i$ are a basis of $h^{1,1}_-$ two-cycles. These do not depend on the string coupling. 

The above string coupling dependence tells us that, at tree-level (and since there are no loop corrections to the perturbative superpotential), the $U_I$ cannot enter the holomorphic Yukawa couplings except through a possible linear term. However, we see that both sets of moduli fields contain axions, and the theory must be invariant under (at least discrete) shifts of $\int_{A_I} C_3 $ and $\int_{C_i}  B_2$. Hence the perturbative contribution to the superpotential must be \emph{independent of  $U_I$}.\footnote{This is only valid as long as there are no background fluxes. In the presence of flux the tree-level superpotential is linear in $U_I$ and up to cubic in $T_i$.} Beyond tree-level, the superpotential can only receive non-perturbative corrections depending on $e^{-2\pi U_I}, e^{-2\pi T_i}$:
\begin{align}
W &= W^{\mathrm{tree}}(e^{-2\pi T_i}) + W^{\mathrm{np}} (e^{-2\pi U_I}, e^{-2\pi T_i}).
\end{align} 
The form for the tree-level superpotential is of course recognised as arising from worldsheet instantons. It can be further restricted when we consider the exact gauging of the axionic symmetries, and how the matter fields transform under them.

For the gauge kinetic function $f_a$ (of a given D6 brane) we know that the tree-level piece, being proportional to the string coupling, must be a linear combination of the complex structure moduli (a linear shift in the gauge kinetic function being permitted as a shift in the theta angle); while the one-loop correction must be independent of the $U_I$ since it must be independent of the string coupling. Hence we can write
\begin{align}
f_a &= \sum_{I=0}^{h^{2,1}} \frac{M_a^I  }{4\pi}U_I  + f_a^{\mathrm{1-loop}} (e^{-2\pi T_i}) + f_a^{\mathrm{np}} (e^{-2\pi U_I}, e^{-2\pi T_i}).
\end{align}

\subsubsection*{IIB}

In type IIB orientifolds with $O3/O7$ planes, the roles of the complex structure and K\"ahler moduli are reversed regarding the string coupling: we have $h^{1,1}$ K\"ahler moduli $T_\alpha$ with 
\begin{align}
T_\alpha \equiv  \frac{1}{\ell_s^6}\int \omega_\alpha \wedge  \bigg( \frac{1}{2 g_s}  J_2 \wedge J_2 - i C_4 \bigg)
\end{align}
where $\omega_\alpha$ is a $(1,1)$-form even under the orientifold projection. Complex structure moduli in type IIB do not have axionic shifts; instead the $H^{2,1}_-$ and $H^{1,2}_-$ forms make up $h^{2,1}_-$ complex scalars $z_I$, independent of the string coupling. Finally the dilaton combines with an R-R zero-form to form the complex scalar $S = \frac{1}{g_s} - i C_0$, and therefore \emph{does} transform linearly under axionic shifts. So in type IIB (for vanishing fluxes) we have
\begin{align}
W &= W^{\mathrm{tree}} (z_I) + W^{\mathrm{np}} (z_I, e^{-2\pi T_i}) \nn\\
f_a &= \frac{k_a S}{4\pi} + \sum_{\alpha=1}^{h^{1,1}_+} \frac{c_a^\alpha  }{4\pi}T_\alpha  + f_a^{\mathrm{1-loop}} (z_I) + f_a^{\mathrm{np}} (z_I, e^{-2\pi S}, e^{-2\pi T_\alpha}).
\label{eq:IIBgeneralform}\end{align}

\subsection{Instanton zero modes and the instanton calculus}
\label{sec:IZM}
To analyse how instantons contribute to the low energy effective action, knowledge of instanton zero modes is absolutely essential. The relevant zero modes correspond to massless excitations of strings with at least one end on the instanton. In path integrals we will have to integrate over these zero modes, which will be the origin of potential new visible sector couplings. In particular, fermionic zero modes play an important role. Unless fermionic zero modes are saturated or projected out the integral over these Grassmann fields gives zero.

Any study of new visible sector couplings due to instantons thus necessary has to begin with an examination of instanton zero modes. A good summary of this topic can be found in the review \cite{09023251} and we will be very brief in the following. In particular, we will mainly discuss which zero mode structure is necessary for the instanton to contribute to the superpotential.
\begin{itemize}
\item \textit{Universal zero modes.} A Euclidean D-brane is always pointlike in the external four dimensions and its position in the external directions gives rise to four universal bosonic zero modes $x^{\mu}$. In addition there will be four universal fermionic zero modes $\theta^{\alpha}$ and $\tau^{\dot{\alpha}}$ corresponding to the supersymmetries broken by the instanton. For so-called O(1) instantons, i.e.~Euclidean branes wrapping a cycle invariant under orientifolding, the two zero modes $\tau^{\dot{\alpha}}$ are projected out. Integrating over the remaining universal zero modes then gives the measure for integrating over superspace $\int \textrm{d}^4 x \int \textrm{d}^2 \theta$. This is the correct measure to give rise to contributions to the superpotential $W$ (see, however, e.g. \cite{Petersson:2007sc}). 
\item \textit{Deformation zero modes.}
There are further bosonic and fermionic zero modes from deformations of the Euclidean brane in internal directions. If present, we would also need to integrate over these, which would not give rise to contributions to the superpotential. As we are interested in corrections to Yukawa couplings in this paper, we only want to consider instantons, where all deformation zero modes are absent. This is the case if the Euclidean brane wraps a rigid cycle. Alternatively, such zero modes can be lifted by fluxes \cite{07080403}.
\item \textit{Charged zero modes.} If the instanton and visible sector brane intersect there will be further zero modes located at the intersection, which will then be charged under the visible sector gauge group. Situations where instanton and visible sector intersect have been studied in detail, as integrating over charged zero modes will typically give rise to new visible sector couplings which are forbidden perturbatively. In contrast, in this paper we study new visible sector couplings due to distant instantons and hence charged zero modes are automatically absent.
\end{itemize}
It follows that superpotential contributions can only be sourced by Euclidean branes where all zero modes except $x^{\mu}$ and $\theta^{\alpha}$ are absent. This is the case for O(1) instantons wrapping a rigid cycle and we take the instantons considered here to be of this type.

\subsubsection*{Stringy instanton calculus}

In string perturbation theory we can directly calculate scattering amplitudes, and extracting superpotential or K\"ahler potential contributions from these can be non-trivial. 
According to the prescription of instanton calculus in string theory (see e.g.~\cite{Billo:2002hm}), to calculate a scattering amplitude in the presence of instantonic branes we must sum over all possible disconnected worldsheets, including matter field and zero mode vertex operators such that all fermionic zero modes are saturated. In general this could be extremely onerous. However, if we want to just extract superpotential contributions, then there are dramatic simplifications due to holomorphy, as we shall briefly review below. Then, to extract the contribution to the superpotential, one option is to calculate scattering amplitudes with only bosons as in \cite{Dine:1986zy,Dine:1987bq} -- probing the scalar potential, leading to effectively instanton--anti-instanton amplitudes or single instanton amplitudes if we know the tree-level superpotential. The (simpler) alternative,  as we shall pursue here, is calculate S-matrix elements involving two fermions and arbitrary numbers of bosons, so that to extract Yukawa couplings we calculate (in Euclidean space): 
\begin{align}
\bra  0| S| \phi_i ,\psi_{j}^\alpha, \psi_k^\beta \ket = - u_{j}^\alpha u_{k}^\beta \frac{e^{\frac{\mathcal{K}}{2}} W_{ijk}}{\sqrt{Z_{i} Z_{j} Z_{k} }}
\label{eq:smatrix}\end{align}
where $W$ is the superpotential, $\mathcal{K}$ the K\"ahler potential, $Z_{i}$ is the (diagonal) K\"ahler metric for the field $i$. The roman indices on the right hand side represent derivatives while the greek ones are spinor indices; $u_j^\alpha$ are spinors. Since the tree-level expressions for the K\"ahler potential and metrics are known, it is therefore straightforward to extract the superpotential term. 
To evaluate the instanton contribution to the left hand side of (\ref{eq:smatrix}) we require the prescription described  in \cite{0609191}:
\begin{itemize}
\item Note that each disk diagram is accompanied by a factor of $1/g_s$, open string loop-diagrams have no factors of $g_s$, and higher-loop amplitudes have positive powers of $g_s$. Since the string coupling $g_s$ corresponds to the dilaton, it appears in the complex combination $S = \frac{1}{g_s} - i C_0$ in the tree-level action where $C_0$ is the axionic zero-form. In addition, in Einstein frame, the string coupling also appears in the expressions for the moduli, in that case linearly accompanied by different axions. In other words, disk diagrams transform by a shift under the axionic symmetries, one-loop diagrams are invariant, but higher-loop diagrams do not have a linear behaviour, and therefore cannot contribute to the general form of (\ref{eq:IIBgeneralform}). Hence only disk and one-loop open-string diagrams contribute. 
\item We must sum over all (disconnected) worldsheets having at least one boundary on the instantonic brane, dividing by symmetry factors for identical worldsheets. Therefore, summing the worldsheets with no vertex operators gives
\begin{align}
\bigg[\sum_{n=0}^\infty \frac{1}{n!}\bigg(\bra 1 \ket_{E}^{\mathrm{disk}} \bigg)^n &\bigg] \times \bigg[ \sum_{n=0}^\infty \frac{1}{n!} \bigg(  \bra 1 \ket_{E,a}^{\mathrm{cylinder}} + \bra 1 \ket_{E,a'}^{\mathrm{cylinder}} + \bra 1 \ket_{E}^{\text{\normalfont m\"obius}}\bigg)^n \bigg] \nn\\
&=  \exp \big[- S_E \big] \times \frac{\mathrm{Pfaff}'(\mathcal{D}_f)}{\sqrt{\mathrm{det}'(\mathcal{D}_B)}},
\label{eq:pfaff}\end{align}
where the pfaffian and determinant factors represent the integration over all of the heavy non-zero modes of the fluctuations of the instanton background. The tree-level action is just $\bra 1 \ket_{Ep} = - \frac{\mu_p}{g_s} V_{p+1} = -\frac{2\pi}{g_s}V_{p+1}/\ell_s^{p+1},$ where $V_{p+1}/\ell_s^{p+1}$ is the size of the instantonic brane in the internal space in units of the string length, $\ell_s$.
\item We must integrate over all zero modes, both fermionic and bosonic, charged and uncharged. The universal \emph{uncharged} bosonic zero modes consist of the four-dimensional position of the instanton -- giving the integration measure $\int \d^4 x$. For fermionic zero modes, in order to have a non-zero amplitude we must have an insertion of each mode. For the universal fermionic zero modes the vertex operators are
\begin{align}
V_{\theta^\alpha}^{-1/2} = \theta^\alpha e^{-\frac{\varphi}{2}} S_\alpha \Sigma_{3/8}^{E,E}
\end{align}
where $\varphi$ is the superconformal ghost, $S_\alpha$ is a left-chiral spin field, which can be bosonised as $e^{\pm \frac{i}{2} (H_4 + H_5)} $ and $\Sigma_{3/8}^{E,E}$ is the spin field on the internal dimensions corresponding to the instanton-instanton open strings. On toroidal orientifolds we have $\Sigma_{3/8}^{E,E} = \prod_{j=1}^3 e^{\frac{i}{2} H_j} $. Note that we require the instanton to be invariant under the orientifold projection, in order to project out additional zero modes with potentially different chirality; since the orientifold action is $\Omega (-1)^{F_L} \sigma$ where $F_L$ counts left-handed spinors in Minkowski space, the $\theta$-modes are invariant only for $O(1)$ instantons (where $\theta^\alpha$ picks up a minus sign from the Chan Paton transformation, rather than plus as for the $U(1)$ case). 
\item The charged zero modes each carry a factor of $\sqrt{g_s}$, and therefore may only appear -- in pairs -- on disk amplitudes (since no other fields carry a factor of $\sqrt{g_s}$). Their  vertex operators are given by
\begin{align}
V_{\lambda_{a,E}}^{-1/2} &= \sqrt{g_s}\lambda_{a,E} e^{-\frac{\varphi}{2}} \sigma_{1/2}^4 \sigma_{1/2}^5  \Sigma_{3/8}^{a,E} \nn\\
V_{\ov{\lambda}_{E,a}}^{-1/2} &= \sqrt{g_s} \ov{\lambda}_{E,a} e^{-\frac{\varphi}{2}} \sigma_{1/2}^4 \sigma_{1/2}^5  \Sigma_{3/8}^{E,a}, 
\end{align}
where the $\sigma_{1/2}^i $ are complex boundary-changing fields with weight $1/8$ for complexified non-compact dimension $i$ (we label Minkowski spacetime as complex dimensions $4$ and $5$).
Note that although the $\lambda, \ov{\lambda}$ are fermions, due to the twist in the spacetime dimensions they carry no four-dimensional spinor indices. Furthermore, while for normal matter fermions the GSO projection relates the orientation of the string with the chirality (distinguishing particle and antiparticle), here only one of $\lambda$ or $\ov{\lambda}$ is kept, depending on how the GSO projection acts on the internal space; we have therefore defined the internal spin fields $\Sigma_{3/8}^{a,E},  \Sigma_{3/8}^{E,a}$ correspondingly. In type IIA, which field we retain depends on whether the intersection number between the brane and instanton is positive or negative. However, we shall not require these fields here, since we are only interested in the effect of distant instantons that do not intersect the visible sector branes.
\item Distributing the matter field and zero mode operators across disks and annuli leads to a potentially large number of diagrams to sum and requires some notation to keep track of them. For the matter fields we introduce a superfield-like notation $\Phi_{a,b}$ which represents either a boson or the combination of a fermion and fermionic zero-mode $\theta$. The indices $a,b$ etc.~indicate the branes that the string ends attach to. We then can construct chains of fields which connect brane $a_k$ to brane $b_k$ via
\begin{align}
\hat{\Phi}_{a_k,b_k} [x_k]&\equiv \Phi_{a_k,x_{k,1}} \Phi_{x_{k,1},x_{k,2}} \cdots \Phi_{x_{k, n_k},b_k}
\end{align}
and therefore if we have $L$ pairs of $\lambda_{a,E}, \ov{\lambda}_{E,b}$ we can denote the collection of disk diagrams as 
$
\prod_{k=1}^L \bra \hat{\Phi}_{a_k,b_k}[x_k] \ket^{\rm disk}_{\lambda_{a_1,E}, \ov{\lambda}_{E,b_1}} .
$
To these we should add annulus diagrams $\prod_k \bra \hat{\Phi}_{c_k,c_k} [x_k]\ket_{c_k,E}^{\rm annulus} $.
\end{itemize}

Putting the above together and including the integration over the zero modes gives
\begin{align}
\nn & \bra  0| S| \phi_i ,\psi_{j}^\alpha, \psi_k^\beta \ket \\
= & - u_{j}^\alpha u_{k}^\beta \mathcal{N}\int \d^4 x \ \d^2 \theta \left(\prod_{\mathrm{uncharged}}d\mu\right) \sum_{\mathrm{configurations}} \prod_a \bigg(\prod_I \d\lambda^I_{a,E} \bigg) \bigg(\prod_J \d\ov{\lambda}^J_{E,a} \bigg) \nn\\
&\times \prod_k \bra \hat{\Phi}_{c_k,c_k} [x_k]\ket_{c_k,E}^{\rm annulus}  \times \prod_{k=1}^L \bra \hat{\Phi}_{a_k,b_k}[x_k] \ket^{\rm disk}_{\lambda_{a_1,E}, \ov{\lambda}_{E,b_1}}  \nn\\
& \times \exp \big[- S_E \big] \times \frac{\mathrm{Pfaff}'(\mathcal{D}_f)}{\sqrt{\mathrm{det}'(\mathcal{D}_B)}}.
\label{eq:whattocalc}\end{align}
Here $\mathcal{N}$ is a normalisation, which we shall give below.
Note that in general we can have additional \emph{uncharged} bosonic or fermionic zero-modes -- where these are present they should be included.

 It was shown in \cite{0612110,Akerblom:2006hx} that the Pfaffian/determinant factors are given by the same expressions as gauge threshold corrections. Recall that gauge threshold corrections can be understood in terms of the Kaplunovsky-Louis formula:
\begin{align}
\frac{1}{g^2_a (\mu)} =& \frac{1}{g_{a\ \rm tree}^2} + \frac{b_0}{16\pi^2}\log \frac{M_s^2}{\mu^2} + \Delta \nn\\
=& \mathrm{Re}(f_a) +  \frac{b_0}{16\pi^2}\log \frac{M_P^2}{\mu^2} + \frac{T_a (G)}{8\pi^2} \log g^{-2}_a + \frac{c_a}{16\pi^2} \mathcal{K} - \sum_R \frac{T_a(R)}{8\pi^2} \log \det K_{R\ov{R}} (\mu).
\label{eq:KaplunovskyLouis}\end{align}
Here $T_a(R) = \mathrm{Tr} (T^2(R))$ for generators $T$ of representation $R$, the sum runs over all fields charged under gauge group $a$, 
\begin{align}
b_0 = \sum_R n_R T_a(R) - 3 T_a (G), \qquad c_a = \sum_R n_R T_a(R) - T_a (G)
\end{align}
for $n_R$ fields in representation $R$, and $K_{R\ov{R}}$ is the K\"ahler metric for matter fields, but we have written the index $R$ to denote summing over representations of the gauge group. In general, showing the equivalence between the first and second line of equation (\ref{eq:KaplunovskyLouis}) is non-trivial; once the K\"ahler metrics have been inserted and the conversion from the (moduli dependent) $M_s$ to $M_P$ (treated as a constant number) is performed, there is typically still a discrepancy which is explained by field redefinitions of the moduli \cite{Akerblom:2007uc,Conlon:2009kt}. Indeed, the calculations in \cite{0612110,Akerblom:2006hx} established the equation on the first line for the instanton case, where we should replace $g_a$ with the gauge coupling that we would have if we replaced the instanton by a brane (or branes) wrapping the same cycle. It was pointed out in \cite{Akerblom:2007uc} that the equivalence with the second line allows us to determine the holomorphic couplings in the superpotential, as follows.
Let us take a general stack of instantons, allow $2L_b$ charged zero modes for each brane $b$, and $n_B, n_F$ uncharged bosonic and fermionic zero modes in total. Then we have 
\begin{align}
c_E = \sum_b 2L_b T_E(R_b) - T_E (G) 
\end{align}
where we use the subscript $E$ to denote the euclidean brane with its fictional gauge group. 
If we consider the disk diagrams, which all have fermionic zero modes inserted, then their supergravity result is 
\begin{align}
\bra \hat{\Phi}_{a_k,b_k}[x_k] \ket^{\rm disk}_{\lambda_{a_1,E}, \ov{\lambda}_{E,b_1}} \equiv \frac{e^{\frac{\mathcal{K}}{2}} Y_{a_k,x_{k,1},...,x_{k,n_k},b_k}}{\sqrt{Z_{\lambda_{a_k,E}} Z_{a_k,x_{k,1}} Z_{x_{k,1},x_{k,2}} \cdots Z_{x_{k,n_k},b_k}Z_{\ov{\lambda}_{E,b_k}} }} 
\end{align}
where $Y_{a_k,x_1,...,x_k,b_k} $ is a holomorphic function with the same moduli dependence as perturbative superpotential terms. Then putting the above together we find
\begin{align}
\bra  0| S| \phi_i ,\psi_{j}^\alpha, \psi_k^\beta \ket& = (- u_{j}^\alpha u_{k}^\beta)\mathcal{N}\int \d^4 x \ \d^2 \theta \left(\prod_{\mathrm{uncharged}}d\mu\right)\sum_{\mathrm{configurations}}  \nn\\
&\times \prod_k \bra \hat{\Phi}_{c_k,c_k} [ x_k]\ket_{c_k,E}^{\rm annulus}  \times \prod_{k=1}^L \frac{Y_{a_k,x_1,...,x_k,b_k}}{\sqrt{ Z_{a_k,x_{k,1}} Z_{x_{k,1},x_{k,2}} \cdots Z_{x_{k,n_k},b_k}}}    \nn\\
& \times \exp \big[- 8\pi^2 f_{E} - T_E(G) \log \mathrm{Re}(f_E)  \big]  e^{T_E(G) \frac{\mathcal{K}}{2}}\nn\\
&\times \exp \big[ - T_E(G) \sum_{\mathrm{uncharged}} \big( \frac{\mathcal{K}}{2} - \log Z_\mu \big) \big] 
\end{align}
By direct computation \cite{Akerblom:2007uc} established that we should assign $T(G)=1$ for $O(1)$ instantons, with this contribution arising from M\"obius strip diagrams. In addition, \cite{0703028,Akerblom:2007uc} gave $\mathcal{N} = \mathrm{Re}(f_E)$ arising from the measure, so that, for $O(1)$ instantons with no additional uncharged zero modes we have
\begin{align}
\bra  0| S| \phi_i ,\psi_{j}^\alpha, \psi_k^\beta \ket =& (- u_{j}^\alpha u_{k}^\beta)\int d^4 x d^2 \theta \sum_{\mathrm{configurations}}  \nn\prod_k \bra \hat{\Phi}_{c_k,c_k} [ x_k]\ket_{c_k,E}^{\rm annulus} \nn\\
& \times \frac{Y_{a_k,x_1,...,x_k,b_k}}{\sqrt{ Z_{a_k,x_{k,1}} Z_{x_{k,1},x_{k,2}} \cdots Z_{x_{k,n_k},b_k}}}  \times \exp \big[- 8\pi^2 f_{E}  \big]  e^{T_E(G) \frac{\mathcal{K}}{2}}, \\
\rightarrow W_{ijk} =&  \int d^4 x d^2 \theta \ e^{- 8\pi^2 f_E}\sum_{\mathrm{configurations}} \prod_k \bra \hat{\Phi}_{c_k,c_k} [\sqrt{Z} x_k]\ket_{c_k,E}^{\rm annulus}  \times \prod_{k=1}^L Y_{a_k,x_1,...,x_k,b_k}  .
\label{eq:penultimatenpW}\end{align}
Here we have 
collected the K\"ahler metric factors and so normalised the annulus amplitudes by multiplying by the K\"ahler metric factors
\begin{align}
\hat{\Phi}_{c_k,c_k} [\sqrt{Z}x_k] \equiv \sqrt{Z_{a_k,x_1}}\Phi_{a_k,x_{k,1}}\sqrt{Z_{x_{k,1},x_{k,2}}} \Phi_{x_{k,1},x_{k,2}} \cdots \sqrt{Z_{x_{k, n_k},c_k}}\Phi_{x_{k, n_k},c_k}.
\end{align}

From the above, the role of disk diagrams to the instanton calculus is clear: they allow new sorts of couplings involving matter fields which can have a very different structure to the perturbative ones (see e.g. \cite{Argurio:2007vqa} as an example in the context of branes at orbifold singularities). However, they require the instanton to intersect the brane. In this work we are interested in the contributions of annulus diagrams. These can contribute whether or not the brane and instanton intersect. In particular, we shall revisit the generation of Yukawa couplings of \cite{0612110} and elucidate some general features of the calculation, but here in the context of type IIB theories with branes at singularities. This will allow us to investigate generation of couplings from instantons distant from the singularity, as in the scenario of \cite{09105496,12071103}. 

\subsection{Role of supersymmetry}
\label{sec:SUSY}

Before beginning explicit calculations, we shall first recall the role of supersymmetry in simplifying the task. Consider that the superpotential contains a term of the form $W \supset e^{-8\pi^2 f_E} \prod_{i=1}^n \Phi_i $ where the $\Phi_i$ are matter fields and $f_E$ depends on both moduli and matter fields. Following the discussion above, we could compute an amplitude involving two fermions and $n-2$ bosons to probe it, saturating the charged and uncharged zero modes. However, firstly such a term in the superpotential will generate several couplings where we permute the particular fermions and bosons; each of them must have the same coefficient by supersymmetry, and so the result should not depend upon which amplitude we calculate. This justifies the notation $\Phi$ as a superfield representing either a boson or fermion; it implies that the amplitudes involving bosons or fermions plus uncharged zero modes must be proportional to each other. Furthermore, in principle we could consider the moduli dependence of the action of the instanton: suppose $f_E$ depends on some modulus $\mathbf{M} = M + \sqrt{2} \theta \tilde{m} + (\theta \theta) F_M + ...$, then the effective Lagrangian should also contain couplings
\begin{align}
\mathcal{L} \supset -8\pi^2 e^{-8\pi^2f_E} \prod_{i=1}^n \phi_i \bigg[  \frac{\partial f_{E}}{\partial \mathbf{M}} F_M +  [ (8\pi^2) \left(\frac{\partial f_{E}}{\partial \mathbf{M}}\right)^2 -   \frac{\partial^2 f_{E}}{\partial \mathbf{M}^2}] \tilde{m} \tilde{m} + ... \bigg] .
\end{align}
If we are considering a modulus with axionic shift symmetries there cannot be a term in the superpotential of the form $e^{-8\pi^2 f_E} \mathbf{M}^p \prod_{i=1}^n \Phi_i $. Furthermore, disk amplitudes involving either two modulus fermions or the operator for the modulus F-term should give exactly the terms in the square brackets. Hence we can compute amplitudes with only matter field bosons (i.e.~no fermions). We will not have to worry about inserting modulus vertex operators onto disks or annuli with matter field operators, because the modulus contribution will have already been taken into account. This has the advantage that the modulus disk amplitudes are universal to all instanton calculations. The proviso is that we can correctly normalise the modulus amplitudes. 

A further implicit assumption in the superfield notation of $\Phi_i$ to represent a boson or fermion is that the equivalent amplitudes involving fermions are always accompanied by uncharged zero modes. This is intuitively correct, in that if we exchange a boson for a fermion in a given amplitude we must also add another fermion in order to have a non-zero result. However, at the CFT level it is somewhat surprising, since although the uncharged zero mode vertex operators correspond to the supersymmetries broken by the instanton, they are placed on a different boundary to the matter fields (the instanton boundary than the visible brane boundary). So it would seem that we cannot simply apply the OPE of the two fields. One of the main results of this paper is to show how this equivalence arises in CFT calculations. 

\subsection{Role of annulus diagrams}

If we calculate an annulus diagram with matter fields, all vertex operators inserted on one boundary must be invariant under all gauge groups, whether anomalous or otherwise. Furthermore, as we have argued above and as we shall show below, we can either calculate the amplitude with two fermions and a number of bosons, or purely with bosons. If we find, say, that an annulus diagram $\bra \prod_{i=1}^m \phi_i\ket_{a,E}^{\rm annulus} \equiv \mathcal{A}_m$ is non-zero, then we should also have contributions from any number of such disjoint annulus diagrams, and thus
\begin{align}
W^{\rm np} &\propto \exp \big[ -8\pi^2 f_E \big] \sum_{n=0}^\infty \frac{1}{n!} \left( - \frac{1}{m!} \mathcal{A}_m \prod_{i=1}^m \Phi_i \right)^n \nn\\
&\propto \exp \big[ -8\pi^2 f_E - \frac{1}{m!}\mathcal{A}_m \prod_{i=1}^m \Phi_i \big].
\end{align}
Hence the role of annulus diagrams is to renormalise the ``gauge kinetic function'' of the instanton, consistent with the expectations of e.g.~\cite{09105496}. We therefore expect that the results for $\mathcal{A}_m$ should be similar to the correction of gauge kinetic functions by matter fields for genuine gauge groups as performed in e.g.~\cite{12071103}, even if the form of the calculation is very different. Our main result in this paper is this calculation for branes at singularities in type IIB. 


From the above, we see that if a combination of fields is gauge invariant for a given value of $m$, then amplitudes for $2m, 3m,\cdots$ fields are likely to be non-zero too. Furthermore, for branes at singularities in type IIB, the matter field vertex operators descend from brane position moduli, which are complex structure moduli;  $f_E$ is a function of these moduli constrained by the symmetries of the compactification. Therefore we can refine equation (\ref{eq:penultimatenpW}) to
\begin{equation}
\boxed{
W^{\rm np} = \exp\big[ - 8\pi^2 f_E - \sum_m \frac{1}{m!} \mathcal{A}_m \prod_{i=1}^m \Phi_i \big]\sum_{\mathrm{configurations}}\prod_{k=1}^L Y_{a_k,x_1,...,x_k,b_k} 
}
\label{eq:npW}\end{equation}
In the absence of intersections between the instanton and visible sector branes, as we will consider in the following calculation, the sum over configurations is replaced by a factor of unity.

\section{String CFT calculation of annulus instanton amplitudes}
\label{sec:calculation}

\subsection{Worldsheet}
\label{sec:worldsheet}
\begin{figure}
	\centering
		\includegraphics[width=0.40\textwidth]{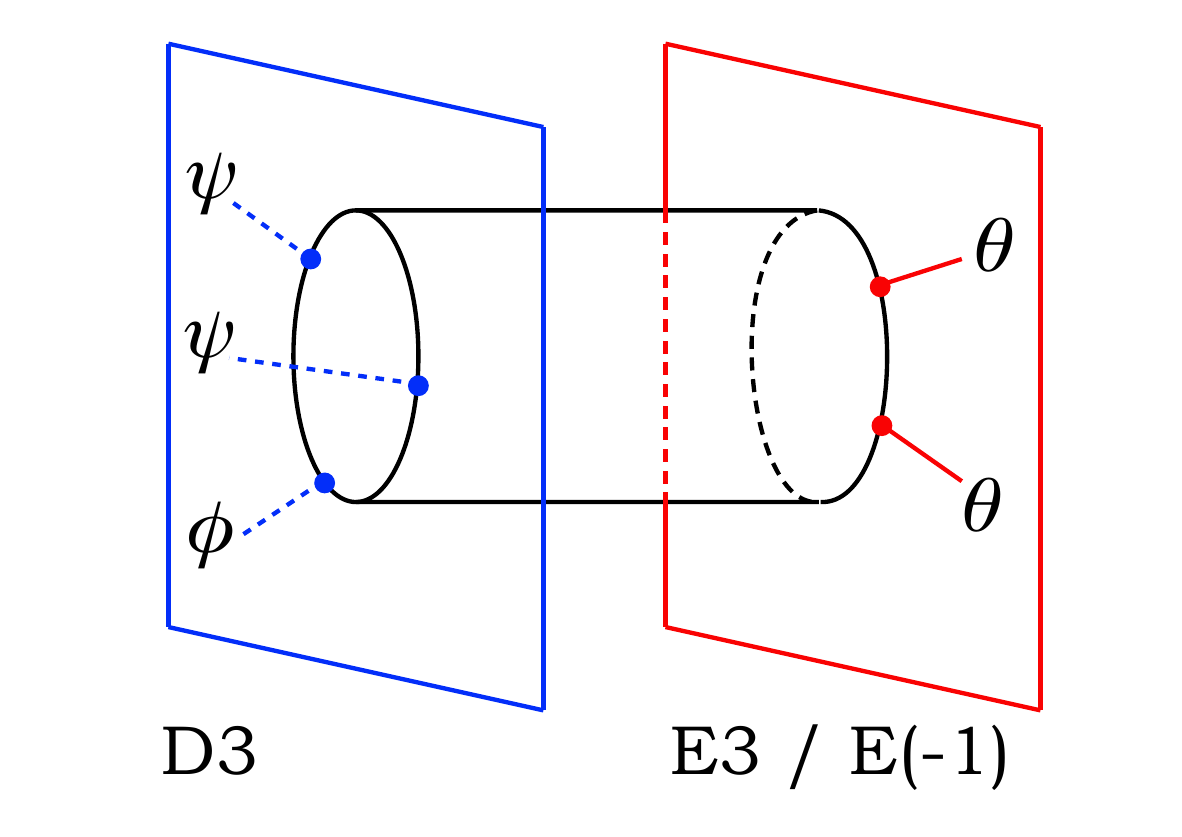}
	\caption{Cylinder worldsheet diagram with one boundary on a stack of D3-branes and the other boundary on a E3 or E(-1) instanton. Vertex operator insertions are given by a visible sector Yukawa operator $\psi \psi \phi$ on the D3 boundary and two universal fermionic zero modes $\theta \theta$ on the instanton boundary. The corresponding amplitude gives information about non-perturbative contributions to Yukawa couplings due to the E3/ E(-1) instanton.}
\label{fig:cylinder}
\end{figure}
To study non-perturbative contributions to Yukawa couplings using CFT techniques, we need to identify the string worldsheets which capture the relevant interactions. Here we are mainly interested in Yukawa couplings due to distant instantons and, correspondingly, any interaction must be mediated by strings stretching between two separated stacks of branes. As a result, relevant worldsheets require at least two boundaries, with one boundary parallel to the visible sector branes and the other boundary aligned with the distant E3/ E(-1) instanton. The lowest order worldsheet with two boundaries in string perturbation theory is the cylinder, which will thus give the leading order contribution to the effect we are considering. In particular, to check for a visible sector Yukawa coupling $\psi_i \psi_j \phi_k$ sourced by an instanton, we place vertex operators corresponding to $\psi_i \psi_j \phi_k$ on one boundary of the cylinder. On the other boundary, we insert vertex operators for two universal instanton zero modes $\theta$. The resulting diagram is shown in figure \ref{fig:cylinder}.

Strictly speaking, our calculations should be performed in an orientifold background. For one, a globally consistent model will typically contain orientifold planes to ensure tadpole cancellation. Further, we invoked orientifolding to project out certain fermionic zero modes. Consequently, we should check how orientifolding could affect our analysis. 
In particular, one could worry whether we also need to examine unoriented worldsheets. However, it is easy to see that such diagrams can only contribute at subleading orders in string perturbation theory. As stated above, new Yukawa couplings due to distant instantons arise from worldsheets with at least two boundaries. To be sensitive to orientifolding the worldsheet also needs to contain at least one cross-cap. As adding a cross-cap introduces an additional factor of $g_s$ any unoriented diagram with two boundaries will necessarily be subleading compared to the cylinder. 

In this paper we wish to examine the generation of new Yukawa couplings in a model-independent way. Rather than performing calculations in a few explicit string theory models, we will thus discuss simplified setups which only retain the phenomenological features necessary for the question at hand. In particular, the constructions considered here will typically not ensure tadpole cancellation. Hence one should check that any non-zero results for the amplitudes evaluated in this work are not artifacts of uncancelled tadpoles. This possibility is easily excluded for the amplitude ${\langle \theta \theta \ \phi_{i} \psi_j \psi_k \rangle}$: tadpoles enter calculations through boundaries without vertex operator insertions. In the diagram considered here (fig.~\ref{fig:cylinder}) we have vertex operator insertions on both cylinder boundaries. Correspondingly the correlator ${\langle \theta \theta \ \phi_{i} \psi_j \psi_k \rangle}$ is not affected by uncancelled tadpoles. However, 1-loop vacuum amplitudes contribute to the Pfaffian and determinant factor in \eqref{eq:whattocalc}. There we carefully remove any dependence on uncancelled tadpoles.

\subsection{Brane setup}
\label{sec:branesetup}
In this work we are interested in non-perturbative contributions from distant Euclidean branes, i.e.~instantons which are spatially separated from the visible sector. In this section we describe the relevant brane setups in toroidal orbifolds. In a full model we should also specify the orientifolding and the corresponding O-planes. However, as discussed in the previous section, unoriented worldsheets do not contribute to the result at leading order. To work with as minimal a setup as possible we will hence not address orientifolding explicitly.

\begin{figure}
	\centering
		\includegraphics[width=0.60\textwidth]{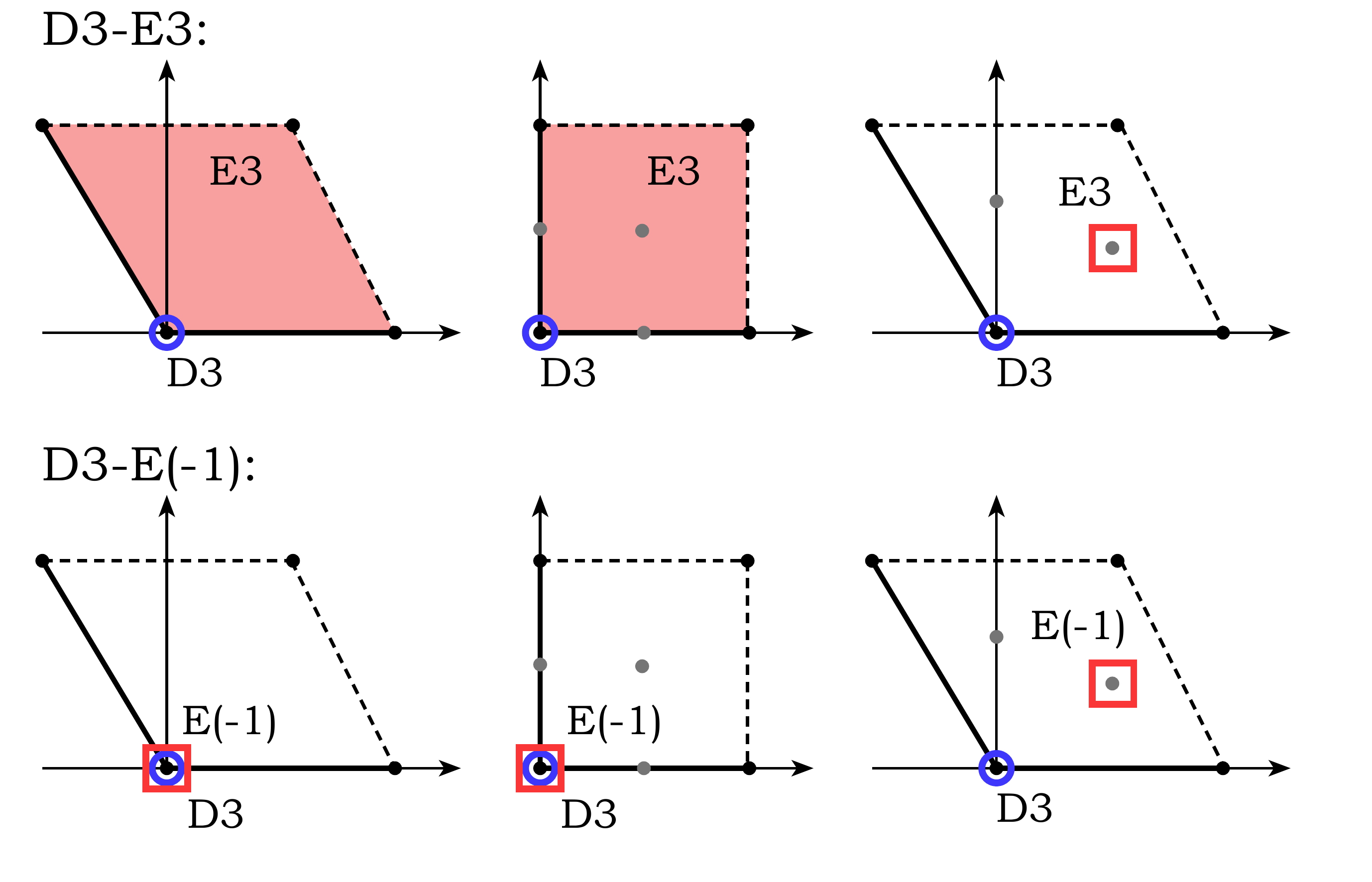}
	\caption{Brane setup for the $\mathbb{T}^6 / \mathbb{Z}_6^{\prime}$ orbifold with geometric orbifold action $\theta = \frac{1}{6} (1,-3,2)$. Black and grey dots denote orbifold fixed points. D3-branes are located at the orbifold singularity at the origin and are marked by a blue circle. In the first setup E3-branes (red) fill out the first two 2-tori completely and are separated from the stack of D3-branes on the third 2-torus. The second configuration contains E(-1)-instantons which are marked by a red square. They are located at an orbifold singularity which is separated from the D3-brane locus along the third 2-torus.}
\label{fig:setup}
\end{figure}

Brane setups of interest are shown in figure \ref{fig:setup} for the case of a $\mathbb{T}^6 / \mathbb{Z}_6^{\prime}$ orbifold. We choose our visible sector to arise from a stack of D3-branes at an orbifold singularity. While not fully realistic the visible sector is sufficient for our considerations, as it exhibits chiral matter and Yukawa couplings. To be specific, we choose the D3-branes to be located at the orbifold singularity at the origin.

There are two types of Euclidean branes which we consider in the following. For one, we will examine E3-branes on bulk cycles. In a toroidal orbifold these correspond to E3-branes wrapping two of the internal 2-tori. Without loss of generality, we choose these to be the first two 2-tori. As we are only interested in setups where the visible sector and instanton branes do not intersect, the E3-branes have to be separated from the D3 stack on the third 2-torus. There, we locate the E3-branes at a singularity distant from that of the D3 (see figure \ref{fig:setup}).

The second class of Euclidean branes we study is given by fractional E(-1) instantons. These are located at orbifold singularities and there are many possibilities of arranging them with respect to the visible sector. As we wish to study effects from distant instantons, the D3-branes and the E(-1) instantons should not coincide. The E(-1) can then be separated from the D3-branes along one, two or all three 2-tori. While we consider all three possibilities, we will shortly see that distant E(-1) instantons can only induce visible sector couplings if they are separated along only one of the 2-tori. To be specific, we choose the E(-1)-instanton to be separated from the D3 stack along the third torus as depicted in figure \ref{fig:setup}. 

We conclude this section with one remark. Non-perturbative effects from E3 instantons play an important role for moduli stabilisation along the lines of KKLT \cite{KKLT} or the Large Volume Scenario (LVS) \cite{LVS}. In the former case the E3-branes wrap a bulk cycle, while in the latter scheme the E3-branes are located on a small blow-up cycle. We are also interested in the question whether the E3 employed for moduli stabilisation can also induce new visible sector couplings. Our brane setups in figure \ref{fig:setup} can be seen as toy models of the situations encountered in KKLT models or the LVS. E3-branes on bulk cycles are modelled by E3-branes wrapping two 2-tori. A fractional E(-1) instanton corresponds to a E3/E1/E(-1) bound state and can thus be seen as an E3 on a blow-up cycle which has collapsed to zero size. While it is not straightforward to generalise results from orbifold calculations to smooth geometries, we can hope to learn more about non-perturbative effects in KKLT and LVS setups by performing a calculation in a toroidal orbifold background. 

\subsection{Winding modes}
\label{sec:winding}
In this paper we calculate instanton induced visible sector couplings by evaluating a cylinder diagram between the branes and the instanton. As we choose the stack of D3s and the instanton to be spatially separated we necessarily require strings stretching across the compact space to connect the two brane stacks. These strings can also wrap the compact space multiple times thus giving rise to a tower of winding states. In the following we will refer to all classical solutions stretching between the visible sector and the instanton as winding solutions. 

Here, we analyse the relevant winding solutions for D3-E3 and D3-E(-1) setups.
Let us parameterise the cylinder worldsheet using the complex coordinates $(z, \bar{z})$, where $\textrm{Re}(z) \in [0, 1/2]$ and $\textrm{Im}(z) \in [0, t/2]$ with $t$ the cylinder modulus. Branes are located at the two cylinder boundaries at $\textrm{Re}(z) =0$ and $\textrm{Re}(z) =1/2$. 

Note that winding solutions only exist for directions with Dirichlet-Dirichlet (DD) boundary conditions at both string endpoints:
\begin{align}
\nn & {\left. \pd_{n} X \right|}_{\textrm{Re}(z) =0} = {\left. \pd_{n} X \right|}_{\textrm{Re}(z) =1/2}=0 \\ 
\Rightarrow \ & X(z, \bar{z}) = x_0 + {\Delta x} \ (z+\bar{z}) + \sum_{oscillators} \ldots \ , \label{eq:winding}
\end{align}
where $x_0$ is the location of one brane stack and $\Delta x$ the separation between the stacks. A Neumann (N) boundary condition $\pd_t X =0$ on either end would be inconsistent with the presence of the term ${\Delta x} \ (z+\bar{z})$. This has the following consequences for our setups of interest.
\begin{enumerate}
\item D3-E3 setups:\newline
In this case we have exactly one internal direction with DD boundary conditions, while the other two internal complex directions exhibit DN boundary conditions. Only one direction gives rise to winding solutions.
\item D3-E(-1): \newline
Here all three internal complex directions have DD boundary conditions and could in principle contribute non-trivial winding modes.
\end{enumerate} 

In addition, there are further boundary conditions on $X$ coming from orbifolding, which will prohibit winding solutions in certain situations. To perform calculations in toroidal orbifolds $\mathbb{T}^6 / \mathbb{Z}_N$ we in practice work with a compact space $\mathbb{T}^6$, but insert an orbifold projection operator $P= \frac{1}{N} (1 + \theta + \ldots \theta^{N-1})$ into the amplitude. As a result, we can split the calculation into different sectors with an orbifold twist of form $\theta^k$ inserted explicitly. The geometric action of $\theta$ can be represented by a vector $\theta=(\theta_1, \theta_2, \theta_3)$, such that
\begin{equation}
\theta^k X^j = e^{2 \pi i k \theta_j} X^j \ , 
\end{equation}
where $X^j$ ($j=1,2,3$) is understood as a complex field. The orbifold sectors fall into three categories:
\begin{itemize}
\item \textit{Fully twisted.} The orbifold action on all subtori is non-trivial: all $\theta_j^k \neq 0$. 
\item \textit{Partially twisted.} The orbifold action leaves one $\mathbb{T}^2$ untwisted: $\theta_j^k = 0$ mod $1$ for exactly one $j$. 
\item \textit{Untwisted sectors.} The orbifold action leaves all tori untwisted.
\end{itemize}

For our cylinder worldsheet the presence of an orbifold twist $\theta^k$ in the amplitude requires that
\begin{equation}
X^j (z+\frac{it}{2}, \bar{z} - \frac{it}{2}) = e^{2 \pi i k \theta_j} X^j(z, \bar{z}) \ .
\end{equation}
For $\theta_j^k \neq 0$ mod 1 the above requires that strings must start and end on the same orbifold fixed point. As a result, for $\theta_j^k \neq 0$ mod 1 we cannot have winding solutions \eqref{eq:winding} with non-zero $\Delta x$. This has the following consequences for setups of interest to us:
\begin{enumerate}
\item D3-E3 setups:\newline
Here the amplitude will receive a contribution from winding modes along the single DD direction in the untwisted sector. In addition, partially twisted sectors are also relevant as long as the untwisted torus coincides with the DD-direction. Fully twisted sectors do not contribute. As every orbifold contains an untwisted sector, winding solutions always exist for E3 instantons on bulk cycles.
\item D3-E(-1) setups:\newline
Here the situation is more complicated and we analyse the possibility of winding solutions sector by sector.
\begin{itemize}
\item \textit{Fully twisted sectors.} The orbifold action on all subtori is non-trivial and hence winding solutions are not allowed. As a result fully twisted sectors do not contribute to the amplitude at all.
\item \textit{Untwisted sectors.} Here we would expect winding solutions for all internal directions. However, while the classical correlator is non-zero, we will find that the quantum amplitude vanishes in the untwisted sector for the D3-E(-1) setup. In consequence, untwisted sectors do not contribute.
\item \textit{Partially twisted sectors.} In this case we can have winding solutions in the one direction which is left untwisted. For the amplitude to be non-vanishing, the D3 and E(-1) branes must then be only be separated along that untwisted direction. This the the only configuration leading to a non-zero amplitude for the D3-E(-1) setup and is shown in figure \ref{fig:setup}. 
\end{itemize}
\end{enumerate} 

\subsection{Pfaffian and determinant factor}
\label{sec:Pfaff}
Following \cite{0609191} the Pfaffian and determinant factor in \eqref{eq:whattocalc} for an O(1) instanton is given by 
\be
\frac{\textrm{Pfaff}^{\prime}(D_F)}{\sqrt{\textrm{det}^{\prime} (D_B)}} = \exp \left( Z^{\prime}(D3,E3) + Z^{\prime}(D3',E3) + Z^{\prime}(E3, O3) \right) \ .
\ee
with a similar expression for the E(-1) case.
Here $Z^{\prime}$ corresponds to 1-loop vacuum amplitudes with open string zero modes removed from the trace and D3$'$ denotes the orientifold image of the visible sector stack of D3-branes. Since D3-branes and E3 instantons are separated in our case, there are no massless open strings in the D3-E3 sector. Correspondingly we can identify $Z^{\prime}(D3,E3)$ with the complete partition function $Z(D3,E3)$.

In the following, we will not evaluate the full Pfaffian and determinant factor. We will only be interested in the part depending on the D3 position, which will arise from the cylinder partition function $Z(D3,E3)$. We begin by evaluating the cylinder partition function for both D3-E3 and D3-E(-1) setups. As mentioned before, only partially twisted sectors contribute in the D3-E(-1) case while both untwisted and partially twisted sectors need to be included for D3-E3 setups. It will be sufficient to only consider expressions in partially twisted sectors. The untwisted case can then be obtained by setting the twist angle $\theta$ to zero. 

Thus we need to evaluate the two following amplitudes
\begin{align}
Z(D3,E3) & = \int_{0}^{\infty} \frac{\textrm{d} t}{t} \sum_{\alpha, \beta =0,1} \delta_{\alpha \beta} \frac{\vartheta_{(\alpha \pm 1) \beta}(0) \vartheta_{(\alpha \pm 1) \beta}(0) \vartheta_{(\alpha \pm 1) \beta}(\theta) \vartheta_{(\alpha \pm 1) \beta}(-\theta)}{\vartheta_4(0) \vartheta_4(0) \vartheta_4(\theta) \vartheta_4(-\theta)} \  \mathcal{Z}_{cl}(r_c,t) \ , \\ 
Z(D3,E(-1)) & = \int_{0}^{\infty} \frac{\textrm{d} t}{t} \sum_{\alpha, \beta =0,1} \delta_{\alpha \beta} \frac{\vartheta_{(\alpha \pm 1) \beta}(0) \vartheta_{(\alpha \pm 1) \beta}(0) \vartheta_{\alpha \beta}(\theta) \vartheta_{\alpha \beta}(-\theta)}{\vartheta_4(0) \vartheta_4(0) \vartheta_1(\theta) \vartheta_1(-\theta)} \  \mathcal{Z}_{cl}(r_c,t) \ ,
\end{align}
where $\mathcal{Z}_{cl}(r_c,t)$ is defined in \eqref{eq:defZcl} and $\delta_{\alpha \beta}=(-1)^{\alpha}$.\footnote{For a detailed discussion of the phases $\delta_{\alpha \beta}$ in the context of cylinder amplitudes involving Euclidean branes see the next section and appendix \ref{sec:sumspin}.} Performing the sum over spin structures using \eqref{eq:Riemann2} and \eqref{eq:Riemann1}, the partition function over string oscillators collapses to a constant in both cases and we are left with
\be
Z(D3,E3) = Z(D3,E(-1)) = \int_{0}^{\infty} \frac{\textrm{d} t}{t} \ \mathcal{Z}_{cl}(r_c,t) \ .
\ee
The above amplitude is sensitive to uncancelled tadpoles, which we need to exclude explicitly. The evaluation is performed in appendix \ref{masspartfuncN2} and gives
\be
Z(D3,E3) = Z(D3,E(-1)) = \mathcal{G}(r_c,\bar{r}_c;0,0) \ ,
\ee
with the Greens function on the torus $\mathcal{G}(w_1, \bar{w}_1; w_2, \bar{w}_2)$ defined as
\begin{equation}
\label{eq:torusG}
\mathcal{G}(w_1, \bar{w}_1; w_2, \bar{w}_2) = - \ln {\left| \frac{\vartheta_1 \left( w_1-w_2, U \right) }{\eta(U)} \right|}^2 + \frac{2 \pi \ [\textrm{Im}(w_1-w_2)]^2}{U_2} \ .
\end{equation}

\subsection{CFT correlator}
\label{sec:CFTcorrelator}
In this section we will evaluate the cylinder correlation function ${\langle \theta \theta \ \phi_{i} \psi_j \psi_k \rangle}$.
We begin by collecting the relevant vertex operators with their canonical ghost charges.
\begin{align}
\nn V_{\phi_3}^{-1} (z_1) & = \hphantom{u_{\alpha}} \ e^{-\varphi} \ e^{i H_3} \ e^{i k_1 \cdot X} (z_1) \\
\nn V_{\psi_3}^{-\frac{1}{2}} (z_2) & = u^{\alpha} \ e^{-\frac{\varphi}{2}} \ S_{\alpha} \ e^{-\frac{i}{2}H_1}e^{-\frac{i}{2}H_2}e^{+\frac{i}{2}H_3} \ e^{i k_2 \cdot X} (z_2) \\ 
\label{eq:vops} V_{\psi_3}^{-\frac{1}{2}} (z_3) & = u^{\beta} \ e^{-\frac{\varphi}{2}} \ S_{\beta} \ e^{-\frac{i}{2}H_1}e^{-\frac{i}{2}H_2}e^{+\frac{i}{2}H_3} \ e^{i k_3 \cdot X} (z_3) \\
\nn V_{\theta}^{-\frac{1}{2}} (z_4) & = \theta^{\gamma} \ e^{-\frac{\varphi}{2}} \ S_{\gamma} \ \prod_{j=1}^3 e^{\frac{i}{2}H_j} (z_4) \\
\nn V_{\theta}^{-\frac{1}{2}} (z_5) & = \theta^{\delta} \ e^{-\frac{\varphi}{2}} \ S_{\delta} \ \prod_{j=1}^3 e^{\frac{i}{2}H_j} (z_5)
\end{align}
where the left-handed spin fields $S_{\alpha}$ can be bosonised as $e^{\pm \frac{i}{2}(H_4 + H_5)}$. Here we chose vertex operators corresponding to a Yukawa coupling $\phi_3 \psi_3 \psi_3$. The reason for this choice will become apparent shortly. The ghost charges sum to (-3) while we require a vanishing ghost charge for a cylinder amplitude. Consequently, we need to picture-change three vertex operators, which we choose to be the three D3-matter field operators. After picture-changing we then have
\begin{align}
\nn V_{\phi_3}^{0} (z_1) & = \hphantom{u_{\alpha} \ e^{+\frac{\varphi}{2}} \ S_{\alpha}} \ \pd Z^3 \ e^{i k_1 \cdot X} (z_1) \\
\nn V_{\psi_3}^{+\frac{1}{2}} (z_2) & = u^{\alpha} \ e^{+\frac{\varphi}{2}} \ S_{\alpha} \ \pd Z^3 \ e^{i k_2 \cdot X} \ \prod_{j=1}^3 e^{-\frac{i}{2}H_j} (z_2) \\ 
V_{\psi_3}^{+\frac{1}{2}} (z_3) & = u^{\beta} \ e^{+\frac{\varphi}{2}} \ S_{\beta} \ \pd Z^3 \ e^{i k_3 \cdot X} \ \prod_{j=1}^3 e^{-\frac{i}{2}H_j} (z_3) \\
\nn V_{\theta}^{-\frac{1}{2}} (z_4) & = \theta^{\gamma} \ e^{-\frac{\varphi}{2}} \ S_{\gamma} \ \prod_{j=1}^3 e^{\frac{i}{2}H_j} (z_4) \\
\nn V_{\theta}^{-\frac{1}{2}} (z_5) & = \theta^{\delta} \ e^{-\frac{\varphi}{2}} \ S_{\delta} \ \prod_{j=1}^3 e^{\frac{i}{2}H_j} (z_5)
\end{align}
where $Z^i$ denote the complexified target space fields defined in \eqref{eq:cplxX}.

The amplitude then factorises into various correlation functions:
\begin{align}
\mathcal{A} = &\sum_{\textrm{cl. sol.}} \int_0^{\infty} \frac{\d t}{t} \int \prod_{i=1}^5 \d z_i \ \langle e^{\frac{\varphi}{2} (z_2)} \nn e^{\frac{\varphi}{2} (z_3)} e^{-\frac{\varphi}{2} (z_4)} e^{-\frac{\varphi}{2} (z_5)} \rangle \ \langle S_{\alpha} (z_2) S_{\beta} (z_3) S_{\gamma} (z_4) S_{\delta} (z_5) \rangle \\
\nn & \langle \prod_{j=1}^3 e^{-\frac{i}{2}H_j (z_2)} \prod_{k=1}^3 e^{-\frac{i}{2}H_k (z_3)} \prod_{l=1}^3 e^{\frac{i}{2}H_l (z_4)} \prod_{m=1}^3 e^{\frac{i}{2}H_m (z_5)} \rangle \ \langle e^{i k_1 \cdot X (z_1)} e^{i k_2 \cdot X (z_2)} e^{i k_3 \cdot X (z_3)} \rangle \\
& \langle \ \pd_n Z^3 (z_1) \ \pd_n Z^3 (z_2) \ \pd_n Z^3 (z_3) \ \rangle \ .
\end{align}
The above amplitude can be evaluated using the standard expressions for correlation functions on the cylinder. Instead of collecting these results in this work, we refer readers to \cite{PolVol1, PolVol2}. Our conventions closely follow \cite{10084361, 12071103}.

\subsubsection*{Bosonic correlation function over $\pd_n Z$}
One important ingredient is the correlator $\langle \pd_n Z^3 \ \pd_n Z^3 \ \pd_n Z^3 \rangle$. The fact that $\pd_n Z^3$ appears can be easily understood. We inserted vertex operators for a Yukawa coupling $\phi_3 \psi_3 \psi_3$, where the subscript indicates that the matter fields arise from fluctuations of the D3-brane in the complex 3-direction. One can easily check that if we inserted vertex operators for a Yukawa coupling $\phi_i \psi_j \psi_k$, we would find 
\begin{equation} 
\label{ZiZjZk}
\langle \ \pd_n Z^i \ \pd_n Z^j \ \pd_n Z^k \ \rangle \subset \mathcal{A} \ .
\end{equation}
Interestingly, only the classical part can contribute to correlators of type \eqref{ZiZjZk},
as the quantum contribution vanishes due to
\begin{equation} 
\pd Z^i (z_1) \pd \bar{Z}^j (z_2) = - \frac{\delta_{ij}}{(z_1-z_2)^2} \ , \qquad \pd Z^i (z_1) \pd Z^j (z_2) = 0 \ .
\end{equation}
Thus, the bosonic correlation function over $\pd_n Z$ will contribute to the overall amplitude as
\begin{equation} 
\label{ZiZjZk2}
\langle \ \pd_n Z^i \ \pd_n Z^j \ \pd_n Z^k \ \rangle = \sum_{\textrm{cl. sol.}} \pd_n Z_{cl}^i \ \pd_n Z_{cl}^j \ \pd_n Z_{cl}^k \ e^{-S_{cl}} \ .
\end{equation}
Non-trivial classical solutions for $Z^i$ are given by winding modes. We discussed the conditions for the existence of such solutions in section \ref{sec:winding}, while explicit expressions can be found in appendix \ref{sec:clsol}. For both D3-E3 and D3-E(-1) setups we found that the result can only depend on winding modes along exactly one internal direction.  WLOG we then choose this direction to be $i=j=k=3$ which corresponds to the choice of vertex operator in \eqref{eq:vops}. To perform the computation we will reexpress \eqref{ZiZjZk2} as in \eqref{eq:pdZclN}.

Intriguingly, already at this stage we find that distant instantons can only induce Yukawa couplings with structure $\phi_i \psi_i \psi_i$, in accordance with the expectation of (\ref{eq:npW}). 

\subsubsection*{Fermionic and $\beta \gamma$-ghost correlators}

To evaluate the fermionic correlators, we find that the non-compact directions yield two amplitudes in each case corresponding to different spinor structures; we can write 
\begin{align}
\bra S_\alpha S_\beta S_\gamma S_\delta \ket =& A_1 \epsilon_{\alpha \beta} \epsilon_{\gamma \delta} + A_2  \epsilon_{\alpha \gamma} \epsilon_{\beta \delta}.
\end{align}
These correspond however to $(\psi \psi) (\theta \theta) $ and $(\psi \theta)(\psi \theta) $. Of course, once we integrate over the fermionic zero modes these will yield $(\psi \psi)$  and $- \frac{1}{2} (\psi \psi)$ respectively. To determine the coefficients $A_{1,2}$ we compute two amplitudes with judicious choices of the external spinors; using $\epsilon_{+-} = +1$ 
\begin{equation}
 A_1 = \bra S_+ S_- S_+ S_- \ket, \qquad A_2 = \bra S_+ S_+ S_- S_- \ket .
\end{equation}
We find that one of the two contributions vanishes after integration over vertex positions, and so we give only the non-zero contribution in the following.

The correlation functions over spinorial fields will differ according to which boundary conditions (Dirichlet vs.~Neumann) are enforced at the cylinder boundaries. Hence we will distinguish between D3-E3 and D3-E(-1) amplitudes in the following. Here and in the following we define $z_{ij} \equiv z_i -z_j$. Our conventions for Jacobi $\vartheta$-functions can be found in appendix \ref{sec:thetaid}.\newline
D3-E3 setup:
\begin{align}
\sum_{\alpha, \beta =0,1} {(-1)}^{\alpha} {\left( \frac{\vartheta_1(z_{25})}{\vartheta_1^{\prime}(0)} \right)}^{-1} & {\left( \frac{\vartheta_1(z_{34})}{\vartheta_1^{\prime}(0)} \right)}^{-1} \ \vartheta_{(\alpha \pm 1) \beta} \left(\frac{z_{25}}{2} - \frac{z_{34}}{2} \right) \ \vartheta_{(\alpha \pm 1) \beta} \left(\frac{z_{25}}{2} - \frac{z_{34}}{2} \right) \ , \\
\nn & \hphantom{Allll} \times \ \vartheta_{(\alpha \pm 1) \beta} \left(- \frac{z_{25}}{2} - \frac{z_{34}}{2} + \theta \right) \ \vartheta_{(\alpha \pm 1) \beta} \left(- \frac{z_{25}}{2} - \frac{z_{34}}{2} - \theta \right) \ .
\end{align}
D3-E(-1) setup:
\begin{align}
\label{eq:D3Dm1prelim}
\sum_{\alpha, \beta =0,1}{(-1)}^{\alpha} \ {\left( \frac{\vartheta_1(z_{25})}{\vartheta_1^{\prime}(0)} \right)}^{-1} {\left( \frac{\vartheta_1(z_{34})}{\vartheta_1^{\prime}(0)} \right)}^{-1} \ & \vartheta_{(\alpha \pm 1) \beta} \left(\frac{z_{25}}{2} - \frac{z_{34}}{2} \right) \ \vartheta_{(\alpha \pm 1) \beta} \left(\frac{z_{25}}{2} - \frac{z_{34}}{2} \right) \\
\nn \times \ & \vartheta_{\alpha \beta} \left(- \frac{z_{25}}{2} - \frac{z_{34}}{2} + \theta \right) \ \vartheta_{\alpha \beta} \left(- \frac{z_{25}}{2} - \frac{z_{34}}{2} - \theta \right)
\end{align}
An important ingredient is the phase $(-1)^{\alpha}$ which ensures the correct summation over spin structures. This phase can be determined by requiring that the amplitude exhibits the correct transformation properties under $z_i \rightarrow z_i + 1$ and $z_i \rightarrow z_i + \tau$, which we show explicitly in appendix \ref{sec:sumspin}. The expression $\delta_{\alpha \beta} = {(-1)}^{\alpha}$ can also be understood as follows. If we replaced the instanton on one boundary of the cylinder by a D-brane wrapping the same internal cycle, the sum over spin structures would need to be performed with the phases $\delta_{\alpha \beta} = {(-1)}^{\alpha + \beta}$ (see e.g.~\cite{12071103}). Recall that the spin structures account for the fact that we need to sum over R and NS sectors while also implementing the GSO projection. However, when deriving the spectrum of Dp-Eq strings, the required GSO projection is exactly opposite to the one for Dp-D(q+4) strings (see e.g.~\cite{0609191, 0703028}). As the GSO projection is implemented by the sum over $\beta$, changing the sign of this projection is equivalent to introducing a phase $(-1)^{\beta}$. It is exactly this factor which accounts for the difference between Dp-Eq and Dp-D(q+4) amplitudes, leading to $\delta_{\alpha \beta} = {(-1)}^{\alpha}$ for Dp-Eq setups.

The sum over spin structures can now be performed using the identity \eqref{eq:Riemann2} for the D3-E3 case and identity \eqref{eq:Riemann1} for D3-E(-1) models.\footnote{At this stage we can confirm explicitly that untwisted sectors cannot contribute to the D3-E(-1) amplitude as asserted in section \ref{sec:winding}: one can check that \eqref{eqD3Dm1check} and hence the whole D3-E(-1) amplitude vanishes for $\theta=0$. On the other hand, the D3-E3 amplitude remains finite for $\theta=0$ as expected.} \newline
\begin{align}
\textrm{D3-E3 setup: } \quad & - 2 \ \frac{{\vartheta_1^{\prime}}^2(0) \ \vartheta_4(-z_{34}) \ \vartheta_4(z_{25}) \ \vartheta_4(\theta) \ \vartheta_4(-\theta)}{\vartheta_1(z_{25}) \ \vartheta_1(z_{34})} \ , \\
\label{eqD3Dm1check} \textrm{D3-E(-1) setup: } \quad & \hphantom{-} 2 \ \frac{{\vartheta_1^{\prime}}^2(0) \ \vartheta_4(-z_{34}) \ \vartheta_4(z_{25}) \ \vartheta_1(\theta) \ \vartheta_1(-\theta)}{\vartheta_1(z_{25}) \ \vartheta_1(z_{34})} \ .
\end{align}

\subsubsection*{Partition functions and correlator over momentum exponentials}
The above correlation functions have to be combined with the partition function over bosonic fields and $bc$-ghosts to arrive at the full quantum amplitude. The partition functions are also sensitive to ND vs.~DD/NN boundary conditions leading to different expressions for the D3-E3 and D3-E(-1) setups:\newline
\begin{align}
\textrm{D3-E3 setup: } \quad & \frac{1}{\vartheta_4(0)} \frac{1}{\vartheta_4(0)} \frac{1}{\vartheta_4 (\theta)} \frac{1}{\vartheta_4 (-\theta)} \frac{1}{\eta^2} \times \eta^2 \ , \\
\textrm{D3-E(-1) setup: } \quad & \frac{1}{\vartheta_4(0)} \frac{1}{\vartheta_4(0)} \frac{(-2 \sin \pi \theta)}{\vartheta_1 (\theta)} \frac{(2 \sin \pi \theta)}{\vartheta_1 (-\theta)} \frac{1}{\eta^2} \times \eta^2 \ ,
\end{align}
where $\eta = \eta(\tau)$ is the Dedekind eta function. The factor $\eta^2$ is the contribution from $bc$-ghosts.
For completeness, we also include the correlator over momentum exponentials
\begin{equation}
\langle e^{i k_1 \cdot X (z_1)} \ e^{i k_2 \cdot X (z_2)} \ e^{i k_3 \cdot X (z_3)} \rangle = \prod_{\substack{i,j=1 \\ i < j}}^3 e^{- k_i \cdot k_j \mathcal{G}(z_{ij})} \ ,
\end{equation}
where $\mathcal{G}(z_{ij})$ is the bosonic propagator for NN boundary conditions. Combining all partial results we arrive at the complete amplitudes.\newline
D3-E3 setup:
\begin{equation}
\label{eq:D3E3a} \mathcal{A}_{D3-E3} = \int_0^{\infty} \frac{\d t}{t} \int \prod_{i=1}^5 \d z_i \ (-2) \ \frac{{\vartheta_1^{\prime}}^2(0) \ \vartheta_4(-z_{34}) \ \vartheta_4(z_{25})}{\vartheta_4(0) \ \vartheta_4(0) \ \vartheta_1(z_{25}) \ \vartheta_1(z_{34})} \ \prod_{\substack{i,j=1 \\ i < j}}^3 e^{- k_i \cdot k_j \mathcal{G}(z_i -z_j)} \langle {(\pd Z^3)}^3\rangle
\end{equation}
D3-E(-1) setup:
\begin{align}
\nn \mathcal{A}_{D3-E(-1)} \ & = \int_0^{\infty} \frac{\d t}{t} \int \prod_{i=1}^5 \d z_i \ (-2) \ {(2 \sin \pi \theta)}^2 \ \frac{{\vartheta_1^{\prime}}^2(0) \ \vartheta_4(-z_{34}) \ \vartheta_4(z_{25})}{\vartheta_4(0) \ \vartheta_4(0) \ \vartheta_1(z_{25}) \ \vartheta_1(z_{34})} \\ \nn & \hphantom{AAAAAAAAAAAll} \times \prod_{\substack{i,j=1 \\ i < j}}^3 e^{- k_i \cdot k_j \mathcal{G}(z_i -z_j)} \langle {(\pd Z^3)}^3\rangle \\ 
\label{eq:D3Dm1a} & = {(2 \sin \pi \theta)}^2 \ \mathcal{A}_{D3-E3}
\end{align}
To extract the contribution of the contact interaction $(\phi \psi \psi) (\theta \theta)$, we only need the momentum-independent part of the amplitude. We thus let $k_i \rightarrow 0$, which does not introduce any subtleties such as momentum poles.\footnote{Momentum poles in the amplitude arise when vertex operators collide ($z_i \rightarrow z_j$) while integrating ${\vartheta_1(z_{ij})}^{-1+ k_i \cdot k_j}$ over worldsheet positions. This does not occur for the amplitudes considered here. Despite the integrand containing factors of the form ${\vartheta_1(z_{25})}^{-1+ k_2 \cdot k_5}$ and ${\vartheta_1(z_{34})}^{-1+ k_3 \cdot k_4}$ the vertex operator positions can never coincide as $z_2, z_3$ are located on one cylinder boundary and $z_4, z_5$ on the other.} As a result, all contributions of the form $e^{-k_i \cdot k_j \mathcal{G}(z_{ij})}$ simply disappear from the amplitude.
In addition, as the amplitudes for both setups are related by a numerical factor, we continue our analysis for $\mathcal{A}_{D3-E3}$ only. The result for $\mathcal{A}_{D3-E(-1)}$ follows immediately from \eqref{eq:D3Dm1a}.

\subsubsection*{Integration over worldsheet positions}
Next, we will perform the integration over vertex operator positions. For the setups at hand all vertex operators correspond to open string excitations and have to be integrated over the cylinder boundaries. Furthermore, the vertex operators for the matter fields $\phi \psi \psi$ are located on one boundary, while the universal zero modes $\theta \theta$ are inserted on the opposite boundary. Thus we have:
\begin{align}
\label{eq:zij1} z_2= \frac{it}{2} y_2 \ , \quad z_3= \frac{it}{2} y_3 \ , \quad z_4=\frac{1}{2} + \frac{it}{2} y_4 \ , \quad z_5=\frac{1}{2} + \frac{it}{2} y_5 \ , \quad \textrm{where } y_i \in [0,1] \ .
\end{align}
For what follows it will be convenient to reexpress the integrand in \eqref{eq:D3E3a} entirely in terms of the theta function $\vartheta_1$. Using the transformation property \eqref{eq:prescription} we rewrite the integrand in \eqref{eq:D3E3a} as:
\begin{align}
\label{eq:4to1} \frac{{\vartheta_1^{\prime}}^2(0) \ \vartheta_4(-z_{34}) \ \vartheta_4(z_{25})}{\vartheta_4(0) \ \vartheta_4(0) \ \vartheta_1(z_{25}) \ \vartheta_1(z_{34})} = e^{\pi i (z_{25} - z_{34})} \ \frac{{\vartheta_1^{\prime}}^2(0) \ \vartheta_1( \frac{it}{4}-z_{34}) \ \vartheta_1( \frac{it}{4} + z_{25})}{\vartheta_1(\frac{it}{4}) \ \vartheta_1(\frac{it}{4}) \ \vartheta_1(z_{25}) \ \vartheta_1(z_{34})}
\end{align}
In addition it will be helpful to transform the amplitude from the ``open string channel'' to the ``closed string channel'', i.e.~write the integrand using theta functions with argument $-\frac{1}{\tau}$ rather than $\tau$. For clarity we reinstate the second argument $\tau= \frac{it}{2}$ of the theta functions $\vartheta_i(z, \tau)$. Applying the transformations \eqref{eq:modtheta1} and \eqref{eq:modtheta1p} to \eqref{eq:4to1} we find:
\begin{align}
\nn & e^{\pi i (z_{25} - z_{34})} \ \frac{{\vartheta_1^{\prime}}^2(0, \frac{it}{2}) \ \vartheta_1( \frac{it}{4}-z_{34},  \frac{it}{2}) \ \vartheta_1( \frac{it}{4} + z_{25},  \frac{it}{2})}{\vartheta_1(\frac{it}{4},  \frac{it}{2}) \ \vartheta_1(\frac{it}{4},  \frac{it}{2}) \ \vartheta_1(z_{25},  \frac{it}{2}) \ \vartheta_1(z_{34},  \frac{it}{2})} \\
= \ & i^2 \ {\left(\frac{2}{t} \right)}^2 \  \frac{{\vartheta_1^{\prime}}^2(0, \frac{2i}{t}) \ \vartheta_1( \frac{1}{2}-\hat{z}_{34}, \frac{2i}{t}) \ \vartheta_1( \frac{1}{2} + \hat{z}_{25}, \frac{2i}{t})}{\vartheta_1(\frac{1}{2}, \frac{2i}{t}) \ \vartheta_1(\frac{1}{2}, \frac{2i}{t}) \ \vartheta_1(\hat{z}_{25}, \frac{2i}{t}) \ \vartheta_1(\hat{z}_{34}, \frac{2i}{t})} \ ,
\end{align}
where we defined $\hat{z} = z / \tau = \frac{2 z}{it}$. Using \eqref{eq:zij1} the coordinates $\hat{z}_{ij}$ are given by
\begin{align}
\label{eq:zij2} \hat{z}_{25} = \frac{i}{t} + y_{25} = i \ell + y_{25} \ , \qquad \hat{z}_{34} = \frac{i}{t} + y_{34} = i \ell + y_{34} \ ,
\end{align}
where $y_{ij} \equiv y_i -y_j$ and we also introduced $\ell = 1 / t$. As a result, in ``closed string channel'' we have
\begin{equation}
\label{eq:D3E3b} \mathcal{A}_{D3-E3} =\frac{(-i)}{4} \int_0^{\infty} \frac{\d \ell}{\ell^4} \int_0^1 \prod_{i=1}^5 \d y_i \ \frac{{\vartheta_1^{\prime}}(0) \ \vartheta_1(\frac{1}{2} -i \ell -y_{34})}{\vartheta_1(\frac{1}{2}) \ \vartheta_1(-i \ell - y_{34})} \ \frac{{\vartheta_1^{\prime}}(0) \ \vartheta_1(\frac{1}{2} +i \ell + y_{25})}{\vartheta_1(\frac{1}{2}) \ \vartheta_1(i \ell + y_{25})}
\ \langle {(\pd Z^3)}^3\rangle \ .
\end{equation}
In the above we also factorised the integrand into two expressions, which are both of the form $\vartheta_1^{\prime}(0) \vartheta_1(a+b) / \vartheta_1(a) \vartheta_1(b)$. To perform the integrals over worldsheet positions, the two factors can be reexpressed using the identity \eqref{eq:qseries}. We give the explicit result for the second factor (involving $y_{25}$):
\begin{align}
\label{eq:y25} \frac{{\vartheta_1^{\prime}}(0) \ \vartheta_1(\frac{1}{2} +i \ell +y_{25})}{\vartheta_1(\frac{1}{2}) \ \vartheta_1(i \ell + y_{25})} = \ & \pi \cot ({\pi}/{2}) + \pi \cot (\pi i \ell  + \pi y_{25}) \\ 
\nn & + 4 \pi \sum_{m,n=1}^{\infty} e^{2 \pi i m n \tau} \ \sin (\pi m +2 \pi i n \ell + 2 \pi n y_{25} ) \ ,
\end{align}
First, note that $\cot(\pi /2) =0$ trivially. It is then straightforward to perform the integrals over $y_i$. In the first step, we change the variables of integration as $y_1, y_2, y_3, y_4, y_5 \rightarrow y_1, y_2, y_{25}, y_{3}, y_{34}$. Note that only $y_{25}$ and $y_{34}$ appear explicitly in \eqref{eq:D3E3b}. In particular, we will need to integrate the expression \eqref{eq:y25} over $y_{25}$. This gives the following result: All contributions involving $\sin$-functions are integrated over entire periods and thus vanish. The only non-zero result comes from the integral over the second $\cot$-term, which is of the form
\begin{align}
\int_{-1}^1 \d y \ \cot(i \epsilon + \pi y) = + 2 i \quad \textrm{for } \epsilon > 0 \ .
\end{align}
The analysis can be repeated for the group of theta functions involving the argument $y_{34}$. Then there is no further explicit dependence on vertex operator positions and the three remaining integrals over $y_i$ give a factor of one. Altogether, integration over worldsheet positions left us with
\begin{align}
\nn \mathcal{A}_{D3-E3} &= (-i) \ \pi^2 \int_0^{\infty} \frac{\d \ell}{\ell^4} \ {\langle \pd Z^3 \pd Z^3 \pd Z^3 \rangle}_{cl} \\ \label{D3E3c} &= (-i) \ \pi^2 \int_0^{\infty} \d t \ t^2 \ {\langle \pd Z^3 \pd Z^3 \pd Z^3 \rangle}_{cl} \ , \\
\label{D3Dm1c} \mathcal{A}_{D3-E(-1)} &= (-i) \ \pi^2 \ {(2 \sin \pi \theta)}^2 \int_0^{\infty} \d t \ t^2 \ {\langle \pd Z^3 \pd Z^3 \pd Z^3 \rangle}_{cl} \ ,
\end{align}
where we also reinstated the sum over classical solutions.\footnote{The amplitudes are now identical to cylinder amplitudes arising in the context of desequestering from gaugino condensation on D7-branes \cite{12071103}.}

\subsubsection*{Integration over worldsheet modulus $t$}
The integral over $t$ can be performed following the analysis in \cite{12071103} which we will sketch briefly. It is convenient to reexpress the classical correlation function as a derivative of the classical partition function w.r.t.~the brane separation as shown in \eqref{eq:pdZclN}:
\begin{align}
\nn {\langle \pd Z^3 \pd Z^3 \pd Z^3 \rangle}_{cl} = \sum_{\textrm{cl.~sol.}} \pd Z_{cl}^3 \pd Z_{cl}^3 \pd Z_{cl}^3 \ e^{-S_{cl}} = {\left(- \frac{\alpha^{\prime}}{t} \sqrt{\frac{2 U_2}{T_2}} \ \partial_{\bar{r}_c} \right)}^3 \ \mathcal{Z}_{cl}(r_c, t) \ .
\end{align}
Then, for example, the D3-E3 amplitude \eqref{D3E3c} can be written as
\begin{equation}
\label{D3E3c2}
\mathcal{A}_{D3-E3} = i \pi^2 \ (\alpha^{\prime})^3 \left( \frac{2 U_2}{T_2} \right)^{3/2} \partial_{\bar{r}_c}^3 \int_0^{\infty} \frac{\d t}{t} \ \mathcal{Z}_{cl}(r_c, t) \ .
\end{equation}
The integral over the classical partition function can be evaluated in terms of the bosonic Greens function $\mathcal{G}(w_1, \bar{w}_1; w_2, \bar{w}_2)$ on a torus defined in \eqref{eq:torusG} (see e.g.~\cite{0404087, 0612110} and appendix \ref{masspartfuncN2}). The final form for the D3-E3 and the D3-E(-1) amplitudes then becomes 
\begin{align}
\label{D3E3d} \mathcal{A}_{D3-E3}  &= - i \pi^2 \ (\alpha^{\prime})^3 \left( \frac{2 U_2}{T_2} \right)^{3/2} \partial_{\bar{r}_c}^3 \ \mathcal{G}(r_c,\bar{r}_c;0,0) \ , \\
\label{D3Dm1d} \mathcal{A}_{D3-E(-1)} &=  -i \pi^2 \ {(2 \sin \pi \theta)}^2 \ (\alpha^{\prime})^3 \left( \frac{2 U_2}{T_2} \right)^{3/2} \partial_{\bar{r}_c}^3 \ \mathcal{G}(r_c,\bar{r}_c;0,0) \ .
\end{align}

\section{Discussion}
\label{sec:discussion}

\subsection{Equivalence of fermionic and bosonic amplitudes}
\label{sec:equivalence}

Here we shall elucidate the general feature of instanton annulus amplitudes expected in section \ref{sec:SUSY}, that the fermionic amplitudes with uncharged zero modes inserted should be equivalent to bosonic amplitudes. To do this, let us look again at 
the integration over the vertex positions of the uncharged zero modes; their coordinates only appear  in the expression given in (\ref{eq:4to1}), so let us define
\begin{align}
f(z_4, z_5) \equiv e^{\pi i (z_{25} - z_{34})} \ \frac{{\vartheta_1^{\prime}}^2(0) \ \vartheta_1( \frac{it}{4}-z_{34}) \ \vartheta_1( \frac{it}{4} + z_{25})}{\vartheta_1(\frac{it}{4}) \ \vartheta_1(\frac{it}{4}) \ \vartheta_1(z_{25}) \ \vartheta_1(z_{34})}.
\end{align}
Note that, as we required above, this function is periodic under shifts in $z_4, z_5$ by $it/2$ but \emph{antiperiodic} under shifts by $z_{4,5} \rightarrow z_{4,5} + 1$, so consider
\begin{align}
\bigg[\int_{1/2}^{1/2+it/2} + \int_{1/2+it/2}^{-1/2+it/2}  + \int_{-1/2+it/2}^{-1/2}   +  \int_{-1/2}^{1/2}  \bigg] \d z_4 f(z_4,z_5)=& 2 \int_{1/2}^{1/2+it/2} \d z_4 f(z_4, z_5)   \nn\\
\rightarrow \int_{1/2}^{1/2+it/2} \d z_4 f(z_4, z_5) = \frac{1}{2} \oint \d z_4 f(z_4, z_5).
\end{align}
Hence we can perform the integrals over $z_4$ and $z_5$ by Cauchy's theorem and find exactly the result that we found before using the theta identities. While this may at first seem like a simple mathematical equivalence, it allows us to generalise the result to \emph{general} backgrounds. 

Let us consider a general annulus instanton amplitude with an uncharged fermionic zero mode. The vertex operator for this field $V_{\theta}^{-1/2}$ having worldsheet coordinate $z$ is $e^{-\phi/2} j_\alpha(z)$, where $j_\alpha$ is the current of the (spacetime) supersymmetry preserved by the D-branes of the theory but broken by the instanton, and $Q_\alpha = \frac{1}{2\pi i}\int e^{-\phi/2} j_\alpha$ is the supercharge. Now, as we argue in appendix \ref{sec:sumspin}, the amplitude must obey (once we apply the doubling trick to extend the annulus to the torus)
\begin{align}
V_{\theta}^{-1/2} (z+1) = - V_{\theta}^{-1/2} (z), \qquad V_{\theta}^{-1/2} (z+it/2) = V_{\theta}^{-1/2} (z)
\end{align}
i.e. the antiperiodicity is required in general. Hence for a generic amplitude  we find
\begin{align}
\bra \cdots V_\theta^{-1/2}\ket_{a,E}^{\mathrm{annulus}} =& \frac{1}{2} \oint dz \bra \cdots e^{-\phi/2} j_\alpha(z)  \ket_{a,E}^{\mathrm{annulus}} 
\end{align}
Crucially, then, we can use the supersymmetry algebra and Cauchy's theorem to extract the amplitude! While the worldsheet realisation of the supersymmetry algebra only closes when we consider all operator pictures together, fortunately we need only consider amplitudes with at most two matter fermions, so we can write all boson operators in the zero picture, matter fields in the $+1/2$ picture and $\theta$ insertions in the $-1/2$ picture. Then the relevant OPEs are (setting the momentum to zero)
\begin{align}
V_\theta^{-1/2}(z) V_\psi^{+1/2}(w) \sim \frac{(\theta u) }{z-w} V_\phi^0 (w) + \mathrm{finite}, \qquad V_\theta^{-1/2}(z) V_\phi^{0}(w) \sim \mathrm{finite}.
\end{align}
These should be true for any theory with $N=1$ supersymmetry.
Therefore we find 
\begin{equation}
\boxed{
\bra V_{\psi_1}^{1/2} V_{\psi_2}^{1/2} V_\theta^{-1/2}  V_\theta^{-1/2} \left(\prod V_\phi^0 \right)\ket_{a,E}^{\mathrm{annulus}} = - 4\pi^2 (\theta u_2) (\theta u_3) \bra  V_{\phi_1}^{0} V_{\phi_2}^{0}\left(\prod V_\phi^0 \right)\ket_{a,E}^{\mathrm{annulus}}
}
\label{eq:equivalence}
\end{equation}
We thus have an explicit demonstration of how spacetime supersymmetry manifests itself in the CFT calculation, and is one of the main results of this paper. We stress that this naively intuitive result, which is expected from \emph{spacetime} supersymmetry, is non-trivial from the worldsheet perspective, in particular since the vertex operators sit at different boundaries and so a priori there is no way that the OPE should apply. 

\subsection{Scaling}
\label{sec:scaling}
We find that both results \eqref{D3E3d} and \eqref{D3Dm1d} for $\mathcal{A}_{D3-E3}$ and $\mathcal{A}_{D3-E(-1)}$
scale as $T_2^{-3/2}$, where $T_2$ is the area of the third 2-torus. In the following we will argue that this behaviour is consistent with supergravity expectations. For simplicity, we will assume that all three tori have the same size $T_2 \sim R^2$. 

By evaluating $\mathcal{A}_{D3-E3}$ and $\mathcal{A}_{D3-E(-1)}$ we extracted information about the effective action. In particular, we calculated the value of the Yukawa coupling $\hat{y}_{ijk}$ in 
\be
\label{eq:yaftercan}
\frac{\hat{y}_{ijk}}{M_s^3} \int \textrm{d}^4 x \ \textrm{d}^2 \theta \ \theta \theta \ \tilde{\phi}^{i} \tilde{\psi}^{j} \tilde{\psi}^{k} = \frac{\hat{y}_{ijk}}{M_s^3} \int \textrm{d}^4 x \ \tilde{\phi}^{i} \tilde{\psi}^{j} \tilde{\psi}^{k}
\ee 
for canonically normalised fields $\tilde{\phi}^{i}$, $\tilde{\psi}^{j}$ and $\tilde{\psi}^{k}$. Our results imply that $\hat{y}_{ijk} \sim T_2^{-3/2} \sim R^{-3}$. 

This should be matched by the supergravity expression. Before canonical normalisation this is given by:
\be
\label{eq:ybeforecan}
\frac{y_{ijk}^{np}}{M_P^3} \int \textrm{d}^4 x \ \textrm{d}^2 \theta \ C^{i} C^{j} C^{k} = \frac{y_{ijk}^{np}}{M_P^3} \int \textrm{d}^4 x \ \phi^{i} \psi^{j} \psi^{k} \ ,
\ee 
Here, $y_{ijk}^{np}$ is the quantity which already appeared in \eqref{eq:npYuk}. It is a perturbative correction to a superpotential term in type IIB string theory and should thus not depend on K\"ahler moduli.

To match expressions \eqref{eq:yaftercan} and \eqref{eq:ybeforecan} we canonically normalise $\tilde{C}^i = \sqrt{K_{i \bar{i}}} C^i$, with $K_{i \bar{i}} \sim \mathcal{V}^{-2/3} \sim R^{-4}$ for matter at a singularity \cite{0609180}. Also recall that Planck scale and string scale are related as $M_P \sim \sqrt{\mathcal{V}} M_s \sim R^3 M_s$. Putting everything together we find 
\be
 \frac{y_{ijk}^{np}}{M_P^3} \int \textrm{d}^4 x \ \phi^{i} \psi^{j} \psi^{k} = \frac{{y}_{ijk}^{np}}{R^3 M_s^3} \int \textrm{d}^4 x \ \tilde{\phi}^{i} \tilde{\psi}^{j} \tilde{\psi}^{k} \quad \Rightarrow \quad \hat{y}_{ijk} = \frac{{y}_{ijk}^{np}}{R^3}
\ee
As ${y}_{ijk}^{np}$ is independent of K\"ahler moduli we hence reproduced the scaling $\hat{y}_{ijk} \sim R^{-3}$. Our result from is thus consistent with supergravity expectations. 

We can now translate our result into an expression for ${y}_{ijk}^{np}$ to find
\be
\label{eq:ycannorm}
{y}_{ijk}^{np} \sim {U_2}^{3/2} \left( \partial_{\bar{r}_c}^3 \ \mathcal{G}(r_c,\bar{r}_c;0,0) \right) e^{\mathcal{G}(r_c,\bar{r}_c;0,0)} \ ,
\ee
where we also included the exponential from the Pfaffian and determinant factor analysed in \ref{sec:Pfaff}. Note that in section in \ref{sec:Pfaff} we did not calculate the terms entering this exponential which do not depend on the brane separation $r_c$. This factor is left implicit in ${y}_{ijk}^{np}$.

\subsection{A numerical result}
\label{sec:num}
We can give a numerical result for the brane position dependent terms entering ${y}_{ijk}^{np}$ in \eqref{eq:ycannorm}. As an example we choose the $\mathbb{T}^6/ \mathbb{Z}_6^{\prime}$ orbifold with $\theta = \frac{1}{6} (1, - 3, 2)$. This corresponds to the setup shown in figure \ref{fig:setup}, where brane and instanton are separated along the third internal direction. The relevant partially twisted sector is generated by $(\theta)^3$ such that $\sin (\pi \theta) = \sin (\pi /2) =1$ in \eqref{D3Dm1d}. Crystallographic restriction on the third 2-torus requires: 
\begin{equation} 
U= - \frac{1}{2} + i \frac{\sqrt{3}}{2} \ , \qquad T_2 =  \frac{\sqrt{3}}{2} R^2 \ .
\end{equation} 
Also, recall that $r_c = \frac{1}{2 \pi} \sqrt{\frac{2 U_2}{T_2}} \ \Delta X$, where $\Delta X$ is the dimensionful complex separation between branes. The definition of the Greens function on the torus $\mathcal{G}$ is given in \eqref{eq:torusG}.

To be specific, let us consider the D3-E(-1) setup. Once we choose to place the D3-branes at the origin, there are two possible loci for the E(-1) which lead to non-zero results. Our findings are summarised in the following table:

\begin{center}
\begin{tabular}{|m{5.5cm}|m{3cm}|m{2.5cm} m{0.1cm}|}
  \hline & setup 1 & setup 2 & \\ \hline \hline
  dim.less D3 locus & $(0,0,0)$ & $(0,0,0)$ & \\[5pt]  \hline
  dim.less E(-1) locus & $(0,0, \frac{i}{\sqrt{3}})$ & $(0,0, \frac{1}{2} + \frac{i}{2 \sqrt{3}})$ & \\[5pt] \hline
$r_c$ & $\frac{i}{\sqrt{3}}$ & $\frac{1}{2} +  \frac{i}{2 \sqrt{3}}$ & \\[5pt]  \hline
 $\left( \partial_{\bar{r}_c}^3 \ \mathcal{G}(r_c,\bar{r}_c;0,0) \right) e^{\mathcal{G}(r_c,\bar{r}_c;0,0)} \approx$ & $16.54 \ i$ & $-16.54 \ i$ & \\[5pt]  \hline
\end{tabular}
\end{center}

One important observation is that $\left( \partial_{\bar{r}_c}^3 \ \mathcal{G}(r_c,\bar{r}_c;0,0) \right) e^{\mathcal{G}(r_c,\bar{r}_c;0,0)} \sim \mathcal{O}(10)$ in the example studied. Thus, despite the spatial separation between brane and instanton, we do not find any numerical suppression of ${y}_{ijk}^{np}$ due to the brane position dependent factors. 

\subsection{Flavour structure}
\label{sec:flavour}
Last, we wish to compare the flavour structure of the non-perturbative contributions to Yukawa couplings with that of the tree level expressions. A stack of $n$ D3-branes at a $\mathbb{Z}_N$ orbifold singularity give rise to a a gauge theory with gauge group \cite{9603167}
\be
\textrm{U}(n_0) \times \textrm{U}(n_1) \times \ldots \textrm{U}(n_{N-1}) \ , \qquad \textrm{with } \sum_{i=0}^{N-1} n_i=n \ .
\ee
The tree level superpotential for chiral matter is inherited from the $\mathcal{N}=4$ superpotential of D3-branes at smooth points. For a $\mathbb{Z}_N$ orbifold with geometric twist $\theta= \frac{1}{N}(v_1, v_2, v_3)$, where $v_1+v_2+v_3 =0$, one finds \cite{9704151}
\be
\label{eq:Wtree}
W_{tree}= \sum_{r,s,t=1}^3 \sum_{i=0}^{N-1} \epsilon_{rst} \ C^{r}_{i, i-v_r} C^{s}_{i-v_r, i-v_r-v_s} C^{t}_{i-v_r-v_s, i} \ .
\ee
Here $\epsilon_{rst}$ is the totally antisymmetric tensor, the superscripts $r,s,t$ label the complex internal direction giving rise to the matter fields $C^r$, $C^s$ and $C^t$. The subscripts $i,j$ on $C^r_{i,j}$ denote the gauge group $\textrm{U}(n_i) \times \textrm{U}(n_j)$ under which the matter field transforms as a bifundamental.

An important observation is that the tree-level Yukawa couplings only contain the field combination $C^1 C^2 C^3$. However, we find that the non-perturbative contributions due to E3/E(-1) instantons always take the form $C^r C^r C^r$. Thus the non-perturbative contributions to Yukawa couplings always involve a different combination of fields than their tree level cousins. For our purposes this is sufficient to conclude that the flavour structure of non-perturbative Yukawa contributions does not align with the flavour structure of the tree level Yukawas.\footnote{Ideally, such a statement should be made in explicit realisations of the (MS)SM in an orbifold setting. However, as we consider the orbifold construction as a toy model for a more realistic compactification on a smooth Calabi-Yau, we do not pursue this point any further.}

For a non-perturbative contribution to Yukawa couplings of the form $C^r C^r C^r$ to be present, this field combination has to be gauge invariant. In the orbifold context the equivalent statement is that the combination $C^r C^r C^r$ has to be invariant under an orbifold twist. Here we want to point out that such situations do occur in toroidal orbifolds of the form $\mathbb{T}^6/\mathbb{Z}_N$. To be specific, let us consider the orbifold $\mathbb{T}^6/\mathbb{Z}_6'$ with orbifold twist $\theta=(\theta_1, \theta_2,\theta_3)=\frac{1}{6}(1,-3,2)$. The group element $\theta^3=\frac{1}{6}(3,-9,6)$ corresponds to a partially twisted sector where the third two-torus is left untwisted. According to the analysis of this paper a Yukawa coupling of the form $C^3 C^3 C^3$ can then be generated by distant E3/E(-1) instantons. We can now check that this combination is indeed gauge invariant. The orbifold acts on the chiral matter fields as $C^r \rightarrow e^{-2 \pi i \theta_r} C^r$. Thus we find 
\be
C^3 C^3 C^3 \rightarrow e^{-2 \pi i (\frac{1}{3}+\frac{1}{3}+\frac{1}{3})} C^3 C^3 C^3=C^3C^3C^3 \ .
\ee
Hence, for this example based on the orbifold $\mathbb{T}^6/\mathbb{Z}_6'$ we conclude that instantons can indeed induce Yukawa couplings of the form $C^3 C^3 C^3$. For an examination of orbifolds beyond $\mathbb{T}^6/\mathbb{Z}_6'$ we refer readers to \cite{12071103}.

\section{Conclusions}
\label{sec:conclusions}
This paper studies non-perturbative contributions to Yukawa couplings in type IIB string theory which are generated by distant Euclidean branes. Calculations are performed in toroidal orbifolds with the visible sector arising from D3-branes at an orbifold singularity. We consider two types of Euclidean branes. For one, we consider E3-branes wrapping a bulk cycle. The second case is given by fractional E(-1) instantons at singularities, which also serve as a toy model for E3 instantons wrapping a blow-up cycle. The existence of new contributions to Yukawa couplings is then checked by evaluating a cylinder worldsheet diagram between the visible sector branes and the instanton with appropriate vertex operator insertions on both boundaries.

Our results can be summarised as follows. A necessary condition for the generation of new visible sector couplings due to distant instantons is the existence of classical (winding) solutions for strings connecting the visible sector with the instanton. For distant E3 instantons wrapping bulk cycles winding solutions connecting the instanton with the visible sector always exist. The full calculation confirms that new visible sector operators are indeed generated which is consistent with previous results \cite{08062291, 09105496}. 

For E(-1) instantons the situation is more complicated. Whether a fractional E(-1) instanton contributes to visible sector couplings depends on the geometrical setup
\begin{itemize} 
\item We find that Yukawa couplings can only be induced if the D3-brane and the E(-1)-instanton are separated along a direction which is left untwisted in a partially twisted sector of the orbifold. 
\item If the orbifold exhibits no partially twisted sector or if the the D3-branes and the fractional E(-1) are separated along a direction which is not left fixed in a partially twisted sector, no visible sector couplings are induced by distant E(-1) instantons.
\end{itemize}
Similar results were obtained in a study of superpotential desequestering due to gaugino condensation on a stack of distant D7-branes \cite{12071103}.

The condition for the generation of Yukawa couplings by fractional E(-1) instantons can also be rephrased in terms of homology. Orbifold singularities which are separated along a direction which is left untwisted in a partially twisted sector exhibit the following: locally all such singularities possess a collapsed 2-cycle, but this 2-cycle is not uniquely associated with a singularity. Rather, all these 2-cycles lie in the same homology class (see e.g.~\cite{11106454}). 

Statements about homology do not only apply to orbifolds, but can also be made for smooth Calabi-Yau compactifications. We can hence conjecture how our results generalise beyond the realm of orbifolds. A straightforward generalisation suggests that distant instantons can induce visible sector couplings if the instanton and the visible sector share a homologous 2-cycle. Hence we expect E3 instantons which satisfy our criterion but do not necessarily wrap bulk cycles also to give rise to new Yukawa couplings.\footnote{Such Yukawa couplings could arise from corrections to E3 superpotential terms due to E1 instantons wrapping the relevant 2-cycle \cite{07103883}. Further investigations in this direction are desirable and we leave them for future work.}

If present, contributions to Yukawa couplings due to E3 instantons take the form
\be
\label{eq:conclYuk}
W \supset Y_{\alpha \beta \gamma}^{np} C^{\alpha} C^{\beta} C^{\gamma} = y_{\alpha \beta \gamma}^{np} C^{\alpha} C^{\beta} C^{\gamma} e^{- a T} \ .
\ee
The cylinder diagram evaluated in this work also gives us direct information about the factor $y_{\alpha \beta \gamma}^{np}$ in \eqref{eq:conclYuk}. In particular, we determine its dependence on the separation between the visible sector and the instanton. Taking the orbifold $\mathbb{T}^6 / \mathbb{Z}_6'$ as an example, we confirm numerically that $y_{\alpha \beta \gamma}^{np}$ is not suppressed despite the separation between the visible sector and the instanton.

Furthermore, we confirm that non-perturbative contributions to Yukawa couplings are potential sources of flavour-violating interactions:
In popular schemes of moduli stabilisation in type IIB string theory, couplings of the form \eqref{eq:conclYuk} give rise to soft A-terms, which inherit their flavour structure: $\delta A_{\alpha \beta \gamma}^{np} \propto y_{\alpha \beta \gamma}^{np}$. For toroidal orbifold models considered here we find that the flavour structure of $y_{\alpha \beta \gamma}^{np}$ and hence of $\delta A_{\alpha \beta \gamma}^{np}$ necessarily differs from that of tree-level Yukawa couplings: $\delta A_{\alpha \beta \gamma}^{np} \neq c Y_{\alpha \beta \gamma}^{tree}$. In this case soft A-terms are possible sources of flavour violation. Avoiding excessive flavour violation will then impose model-dependent constraints on the string theory setup, which are expected to be most restrictive on sequestered realisations of the Large Volume Scenario \cite{10121858}.

In addition, our results are also useful from a technical point of view. So far, the study of 1-loop diagrams involving Euclidean branes in type IIB string theory was limited to vacuum amplitudes, while calculations with vertex operator insertions were only performed in type IIA string theory \cite{0612110}. Here we performed a similar calculation with vertex operator insertions on the type IIB side, where important steps such as the summation over spin structures are described in detail. In doing so, we have elucidated the role of annulus diagrams with matter fields for instanton-generated superpotentials, our main conclusions being summarised in equations (\ref{eq:npW}) and (\ref{eq:equivalence}). In particular, 
we proved how spacetime supersymmetry manifests itself in terms of the equivalence between annulus diagrams with bosons, fermions and uncharged zero modes inserted, and those with only bosons. 

We hope that the techniques employed in this work will prove useful for future calculations of 1-loop diagrams in this context. In particular, the results leading to equation (\ref{eq:equivalence}) can also be used to prove identities involving additional uncharged zero modes on  instantons, such as those suggested in \cite{Blumenhagen:2008ji}. 

An important open question is the generalisation of these results to smooth compactifications on Calabi-Yau orientifolds, perhaps using techniques along the lines of \cite{09105496}. In particular, it would be interesting to examine whether our condition in terms of homology continues to hold.

\section*{Acknowledgments}
We thank Ralph Blumenhagen, Joseph Conlon, Arthur Hebecker, M.C.~David Marsh, Eran Palti and Timo Weigand for insightful discussions and helpful comments. We are also grateful to Arthur Hebecker for comments on the draft of this paper. LW acknowledges support by the DFG Transregional Collaborative Research Centre TRR 33 ``The Dark Universe''. We would like to thank the organisers of IFT workshop ``New Challenges in String Phenomenology'' in Madrid for hospitality. MDG thanks the Institut Lagrange de Paris.

\appendix

\section{Theta functions and identities}
\label{sec:thetaid}
\subsubsection*{Definitions}
The theta function with characteristics is defined as
\begin{equation}
\vartheta \left[\begin{array}{c} a \\ b \end{array}\right] \left(z, \tau \right) = \sum_n e^{\pi i (n+a)^2 \tau + 2 \pi i (n+a) (z+b)} \ .
\end{equation}
We will also use the notation $\vartheta_{\alpha \beta}(z, \tau) \equiv \tha{ \alpha/2 }{ \beta/2}{z, \tau}$ as well as
\begin{equation}
\vartheta_1 \equiv \vartheta_{11} \ , \qquad \vartheta_2 \equiv \vartheta_{10} \ , \qquad \vartheta_3 \equiv \vartheta_{00} \ , \qquad  \vartheta_4 \equiv \vartheta_{01} \ .
\end{equation}
In calculations we will typically suppress the second argument: $\vartheta_{\alpha \beta} (z, \tau) = \vartheta_{\alpha \beta} (z)$.

\subsubsection*{Identities}
As long as $|\textrm{Im}(a)| < \textrm{Im}(\tau)$ and $|\textrm{Im}(b)| < \textrm{Im}(\tau)$ one has
\begin{align}
\label{eq:qseries} \frac{\vartheta_1^{\prime}(0) \ \vartheta_1(a+b)}{\vartheta_1(a) \ \vartheta_1(b)} = & \ \pi \cot (\pi a) + \pi \cot (\pi b) + 4 \pi \sum_{m,n=1}^{\infty} e^{2 \pi i m n \tau} \ \sin (2 \pi m a + 2 \pi n b) \ ,
\end{align}
where here, and in what follows, $\vartheta_1^{\prime}(0) = \frac{\d}{\d z} \vartheta_1(z) |_{z=0}$.

\subsubsection*{Riemann Theta formulae}
Theta functions exhibit a multitude of identities involving sums over products of four theta functions (see e.g.~\cite{Mumford1}). We will perform sums over spin structures using the following identities:
\begin{align}
\nn &- \vartheta_{00}(x) \vartheta_{00}(y) \vartheta_{00}(u) \vartheta_{00}(v) - \vartheta_{01}(x) \vartheta_{01}(y) \vartheta_{01}(u) \vartheta_{01}(v) \\ \nn  & + \vartheta_{10}(x) \vartheta_{10}(y) \vartheta_{10}(u) \vartheta_{10}(v) + \vartheta_{11}(x) \vartheta_{11}(y) \vartheta_{11}(u) \vartheta_{11}(v) \\
\label{eq:Riemann2} = -2 \ & \vartheta_{01}(\frac{x+y+u+v}{2}) \vartheta_{01}(\frac{x+y-u-v}{2}) \vartheta_{01}(\frac{x-y+u-v}{2}) \vartheta_{01}(\frac{x-y-u+v}{2}) \ ,
\end{align}
and
\begin{align}
\nn &\vartheta_{00}(x) \vartheta_{00}(y) \vartheta_{00}(u) \vartheta_{00}(v) + \vartheta_{01}(x) \vartheta_{01}(y) \vartheta_{01}(u) \vartheta_{01}(v) \\ \nn - & \vartheta_{10}(x) \vartheta_{10}(y) \vartheta_{10}(u) \vartheta_{10}(v) - \vartheta_{11}(x) \vartheta_{11}(y) \vartheta_{11}(u) \vartheta_{11}(v) \\
\label{eq:Riemann1} = 2 \ & \vartheta_{01}(\frac{x+y+u+v}{2}) \vartheta_{01}(\frac{x+y-u-v}{2}) \vartheta_{11}(\frac{x-y+u-v}{2}) \vartheta_{11}(\frac{x-y-u+v}{2}) \ .
\end{align}

\subsubsection*{Transformations under shifts of argument}
\begin{align}
\label{minus1}
\vartheta \left[\begin{array}{c} a \\ b \end{array}\right] \left(z-1, \tau \right) \ & =
e^{-2 \pi i a} \vartheta \left[\begin{array}{c} a \\ b \end{array}\right] \left(z, \tau \right) \\ 
\label{minustau} \vartheta \left[\begin{array}{c} a \\ b \end{array}\right] \left(z-\tau, \tau \right) \ & =
e^{2 \pi i b} e^{- \pi i \tau} e^{2 \pi i z} \vartheta \left[\begin{array}{c} a \\ b \end{array}\right] \left(z, \tau \right) \\ 
\label{plushalf} \vartheta \left[\begin{array}{c} a \\ b \end{array}\right] \left(z+\frac{1}{2}, \tau \right) \ & = 
\begin{cases} \hphantom{(-1)} \  \vartheta \left[\begin{array}{c} a \\ 1/2 \end{array}\right] \left(z, \tau \right) \qquad b=0 \\ 
\hphantom{(-1)} \  \vartheta \left[\begin{array}{c} 0 \\ 0 \end{array}\right] \left(z, \tau \right) \qquad \hphantom{lll} b=1/2 \quad a=0 \\ 
{(-1)} \ \vartheta \left[\begin{array}{c} 1/2 \\ 0 \end{array}\right] \left(z, \tau \right) \qquad b=1/2 \quad a=1/2 \end{cases} \\ 
\label{plushalftau} \vartheta \left[\begin{array}{c} a \\ b \end{array}\right] \left(z+\frac{\tau}{2}, \tau \right) \ & =
e^{- \pi i b} e^{- \frac{\pi i}{4} \tau} e^{- \pi i z} \vartheta \left[\begin{array}{c} a - 1/2 \\ b \end{array}\right] \left(z, \tau \right) 
\end{align}
One special case of \eqref{plushalftau} which will be particularly useful is
\begin{equation}
\label{eq:prescription}
\vartheta_{01}(z) = i \ e^{\frac{\pi i}{4} \tau} \ e^{\pi i z} \ \vartheta_{11} \left(z + \frac{\tau}{2} \right)
\end{equation}

\subsubsection*{Modular transformations}
To transform cylinder amplitudes from the ``open string channel'' to the ``closed string channel'' we will require
\begin{align}
\label{eq:modtheta1} \vartheta_1(z, \tau) & = i (-i \tau)^{-1/2} e^{- \pi i z^2 / \tau} \ \vartheta_1 \left( \frac{z}{\tau}, -\frac{1}{\tau} \right) \ , \\
\label{eq:modtheta1p} \vartheta_1^{\prime}(0, \tau) & = \hphantom{i} (-i \tau)^{-3/2} \ \vartheta_1^{\prime} \left( 0, -\frac{1}{\tau} \right) \ .
\end{align}

\section{Classical solutions}
\label{sec:clsol}
Here we will collect explicit expressions for winding solutions for later use. Let the three internal 2-tori be parameterised by the coordinates $X^{M}$ with $M=4, \ldots, 9$ subject to the identification $X^M \sim X^M+1$. 
We can then define complex coordinates $Z^i$ and $\bar{Z}^i$ as in \cite{0404134}:
\begin{equation}
\label{eq:cplxX}
Z^i = \sqrt{\frac{T_2^i}{2 U_2^i}} (X^{2i+2} + \bar{U}^i X^{2i+3}) \ , \qquad \bar{Z}^i = \sqrt{\frac{T_2^i}{2 U_2^i}} (X^{2i+2} + {U}^i X^{2i+3}) \ ,
\end{equation}
where $U^i$ are the complex structure moduli and $T_2^i$ parameterise the volumes of the 2-tori.

We now want to write down the classical solution $Z_{cl}^i$ for a string starting at a brane at position $z_0^i$ and ending at a brane at $z_0^i + \Delta z^i$. The cylinder worldsheet with coordinates $(w, \bar{w})$ is embedded in such a way that $\Re (w) =0$ corresponds to the locus of one brane, while the other brane is placed at $\Re(w) = 1/2$. We then have
\begin{equation}
Z_{cl}^i(w, \bar{w}) = z_0^i + 2 \pi \sqrt{\frac{T_2^i}{2 U_2^i}} (m + \bar{U}^i n + r_c)(w + \bar{w}) \ ,
\end{equation}
where $m,n$ are integers and we defined the dimensionless brane separation $r_c = \frac{1}{2 \pi} \sqrt{\frac{2 U_2^i}{T_2^i}} \ \Delta z^i$. The integers $m,n$ correspond to different winding states. In this paper we will need the classical solution $\partial_n Z_{cl}^i$ for strings stretching between separated branes, where $\partial_n$ is a directional derivative normal to the brane. This is easily obtained from the above expression:
\begin{equation}
\pd_n \ Z_{cl}^i(w, \bar{w}) = \frac{1}{2} (\pd_w + \pd_{\bar{w}}) \ Z_{cl}^i(w, \bar{w})= 2 \pi \sqrt{\frac{T_2^i}{2 U_2^i}} (m + \bar{U}^i n + r_c) \ .
\end{equation}

Furthermore, the classical solutions will also enter our results via the exponential of the classical action, with which we will need to weigh the correlation function. For one internal complex direction we get (see e.g.~\cite{10084361})
\begin{equation}
\label{eq:defZcl}
\mathcal{Z}_{cl}(r_c,t) \equiv \sum_{\textrm{cl. sol.}} e^{- S_{cl}} = \sum_{m,n=-\infty}^\infty \exp \left(- \frac{2 \pi \ t}{\ap} \frac{T_2^i}{2 U_2^i} |m + \bar{U}^i n + r_c|^2 \right) \ ,
\end{equation}
with $t$ the modulus of the cylinder worldsheet.

One result which will be useful later on is the classical correlation function (for the $i$-th complex direction) for $N$ insertions of $\pd Z^i$:
\begin{align}
\nn \langle \ ( \pd_n Z^i )^N \rangle_{cl} \ &=
\sum_{\textrm{cl.~sol.}} ( \pd_n Z_{cl}^i )^N \ e^{-S_{cl}} \\
\nn &= \sum_{m,n=-\infty}^\infty {\left[ (2 \pi) {\left( \frac{T_2^i}{2 U_2^i} \right)}^{1/2} (m + \bar{U}^i n + r_c) \right]}^N \ e^{- \frac{2 \pi t}{\ap} \frac{T_2^i}{2 U_2^i} |m + \bar{U}^i n + r_c|^2} \\
\nn &= {\left(- \frac{\alpha^{\prime}}{t} \sqrt{\frac{2 U_2^i}{T_2^i}} \ \partial_{\bar{r}_c} \right)}^N  \sum_{m,n=-\infty}^{\infty} e^{- \frac{2 \pi t}{\ap} \frac{T_2^i}{2 U_2^i} |m + \bar{U}^i n + r_c|^2} \\
\label{eq:pdZclN} &= {\left(- \frac{\alpha^{\prime}}{t} \sqrt{\frac{2 U_2}{T_2}} \ \partial_{\bar{r}_c} \right)}^N \ \mathcal{Z}_{cl}(r_c, t) \ ,
\end{align}
which is now conveniently expressed as a derivative of the classical partition function w.r.t.~the brane separation.

\section{Spinor part of the CFT amplitude}
\label{sec:sumspin}
In this section we demonstrate how to correctly perform the sum over spin structures for the amplitudes considered in this work. To be specific, we will use the D3-E(-1) desequestering amplitude as an example. The spin-structure dependent part of the amplitude arises from correlation functions over fermionic fields and ghosts, which for the D3-E(-1) setup is given by (see also \eqref{eq:D3Dm1prelim}) 
\begin{align}
\label{eq:D3Dm1prelim2}
\sum_{\alpha, \beta =0,1} \delta_{\alpha \beta} \ {\left( \frac{\vartheta_1(z_{25})}{\vartheta_1^{\prime}(0)} \right)}^{-1} {\left( \frac{\vartheta_1(z_{34})}{\vartheta_1^{\prime}(0)} \right)}^{-1} \ & \vartheta_{(\alpha \pm 1) \beta} \left(\frac{z_{25}}{2} - \frac{z_{34}}{2} \right) \ \vartheta_{(\alpha \pm 1) \beta} \left(\frac{z_{25}}{2} - \frac{z_{34}}{2} \right) \\
\nn \times \ & \vartheta_{\alpha \beta} \left(- \frac{z_{25}}{2} - \frac{z_{34}}{2} + \theta \right) \ \vartheta_{\alpha \beta} \left(- \frac{z_{25}}{2} - \frac{z_{34}}{2} - \theta \right) \ ,
\end{align}
At the moment the phases $\delta_{\alpha \beta}$ are left undetermined in the above. In the following we will show how the phases $\delta_{\alpha \beta}$ can be determined up to an overall phase from requiring that the amplitude exhibits the expected behaviour under shifts of vertex operator positions $z_i \rightarrow z_i + 1$ and $z_i \rightarrow z_i + \tau$.

\subsection*{Behaviour of $\mathcal{A}$ under shifts $z_i \rightarrow z_i + 1$ and $z_i \rightarrow z_i + \tau$}
It will be enough to consider the behaviour under the shift of one vertex operator, which WLOG we choose to be 
\be
\label{eq:refop}
V_{\theta}^{-\frac{1}{2}} (z_4) = \theta^{\alpha} \ e^{-\frac{\varphi}{2}} \ S_{\alpha} \ \prod_{j=1}^3 e^{\frac{i}{2}H_j} (z_4) \ .
\ee

\begin{itemize}
\item Shift $z_4 \rightarrow z_4 + \tau$:\newline
The cylinder amplitude should be periodic up to phases due to the orbifold twist. In particular, under an orbifold twist the above vertex operator \eqref{eq:refop} changes as
\be
V_{\theta}^{-\frac{1}{2}} \rightarrow e^{\frac{\pi i}{2} \theta_1 + \frac{\pi i}{2} \theta_2 + \frac{\pi i}{2} \theta_3} \ V_{\theta}^{-\frac{1}{2}} \ .
\ee
Here, all calculations are performed in partially twisted sectors with $\theta_1=-\theta_2=\theta$ and $\theta_3=0$. We thus find 
\be
V_{\theta}^{-\frac{1}{2}} \rightarrow e^{\frac{\pi i}{2} \theta - \frac{\pi i}{2} \theta} \ V_{\theta}^{-\frac{1}{2}} = V_{\theta}^{-\frac{1}{2}} \ .
\ee
As a result, the amplitude is periodic under shifts $z_4 \rightarrow z_4 + \tau$.
\item Shift $z_4 \rightarrow z_4 + 1$:\newline
The behaviour under shifts $z_4 \rightarrow z_4 + 1$ will depend on the boundary conditions imposed by the branes. In the following, we will closely follow the discussion in \cite{0211250}. To begin, note that the D3-branes and the E(-1) instanton preserve different supersymmetry charges. We can write the left and right-moving charges in ten dimensions as 
\be
Q^{\dot{\mathcal{A}}} = \frac{1}{2 \pi i} \int \textrm{d} z \ e^{- \frac{\varphi}{2}} S^{\dot{\mathcal{A}}} (z) \ , \qquad \tilde{Q}^{\dot{\mathcal{A}}} = \frac{1}{2 \pi i} \int \textrm{d} \zb \ e^{- \frac{\tilde{\varphi}}{2}} \tilde{S}^{\dot{\mathcal{A}}} (\zb) \ ,
\ee
where $S^{\dot{\mathcal{A}}}$, $\tilde{S}^{\dot{\mathcal{A}}}$ are 10d spin fields and $\dot{\mathcal{A}}$ is a 10d antichiral spinor index. The branes preserve a combination of left and right moving charges given by
\be
Q^{\dot{\mathcal{A}}} + \prod_{m} \beta^m \tilde{Q}^{\dot{\mathcal{A}}} \ ,
\ee
where $m$ runs over all directions perpendicular to the brane and $\beta^m \equiv \Gamma \Gamma^m$, where $\Gamma^m$ are the 10d Dirac matrices and $\Gamma \equiv \Gamma^0 \Gamma^1 \ldots \Gamma^9$.

In the following it will be useful to decompose the 10d spin fields as products of 4d and 6d spin fields: $S^{\dot{\mathcal{A}}} \rightarrow (S_{\alpha} S_A, S^{\dot{\alpha}} S^A)$. The conditions for the preservation of supersymmetry can then be rewritten as conditions on the spin fields on the cylinder boundaries. Here we have placed the D3-brane at $\textrm{Re(z)}=0$ while the E(-1) instanton is located at $\textrm{Re(z)}=1/2$. We then have
\begin{align}
\label{eq:Dm1susycond} E(-1): \quad S_{\alpha} S_A (z) &=  \left. \epsilon \tilde{S}_{\alpha} \tilde{S}_A (\zb) \right|_{z=1-\zb} \ , \quad &S^{\dot{\alpha}} S^A (z) &= \left. \hphantom{-} \epsilon \tilde{S}^{\dot{\alpha}} \tilde{S}^A (\zb) \right|_{z=1-\zb} \ , \\
\label{eq:D3susycond} D3: \quad S_{\alpha} S_A (z) &=  \left. \epsilon'  \tilde{S}_{\alpha} \tilde{S}_A (\zb) \right|_{z=-\zb} \ , \quad  &S^{\dot{\alpha}} S^A (z) &= \left. - \epsilon' \tilde{S}^{\dot{\alpha}} \tilde{S}^A (\zb) \right|_{z=-\zb} \ , 
\end{align}
where $\epsilon=\pm 1$, $\epsilon'=\pm 1$ depending on whether we are considering branes or antibranes. As only the relative sign $\epsilon \epsilon'$ has physical significance, WLOG we can set $\epsilon=-1$.

Now we wish to determine the remaining sign $\epsilon'$ and deduce the behaviour of vertex operators under shifts $z_i \rightarrow z_i+1$. To this end notice that $V_{\theta}^{-\frac{1}{2}} = e^{- \frac{\varphi}{2}} S_{\alpha} S_A$. Furthermore, the fermionic zero mode $\theta$ corresponds to a goldstone field, which is associated to a supersymmetry broken by the E(-1), but not by the D3-brane. Thus $V_{\theta}^{-\frac{1}{2}}$ should be associated to a supersymmetry charge broken by the E(-1) but preserved by the D3.

To this end consider the charge $Q_{\alpha A} + \tilde{Q}_{\alpha A}$. Writing $z=x+iy$ the condition \eqref{eq:Dm1susycond} implies: 
\begin{align}
\nn E(-1): \quad \left. Q_{\alpha A} + \tilde{Q}_{\alpha A} \right|_{z=1-\zb} &= \frac{1}{2 \pi i} \int \left. \left(\d z \ e^{- \frac{\varphi}{2}} S_{\alpha} S_A +\d \zb \ e^{- \frac{\tilde{\varphi}}{2}} \tilde{S}_{\alpha} \tilde{S}_A  \right) \right|_{z=1-\zb} \\ 
\nn &= \frac{1}{2 \pi i} \int \left( i \d y \ e^{- \frac{\varphi}{2}} S_{\alpha} S_A - (-i) \d y \ e^{- \frac{\varphi}{2}} S_{\alpha} S_A \right) \\
\label{eq:brokenbyDm1} &= \frac{1}{\pi} \int \d y \ e^{- \frac{\varphi}{2}} S_{\alpha} S_A = \frac{1}{\pi} \int \d y \ V_{\theta}^{-\frac{1}{2}} \ .
\end{align}
This shows that $V_{\theta}^{-\frac{1}{2}}$ indeed corresponds to a goldstino of a supersymmetry broken by E(-1). To be consistent with our interpretation of $V_{\theta}^{-\frac{1}{2}}$ as the vertex operator for the universal zero mode $\theta$, the charge $Q_{\alpha A} + \tilde{Q}_{\alpha A}$ has to be preserved by the D3. Using \eqref{eq:D3susycond} we find
\begin{align}
\nn D3: \quad \left. Q_{\alpha A} + \tilde{Q}_{\alpha A} \right|_{z=-\zb} &= \frac{1}{2 \pi i} \int \left. \left(\d z \ e^{- \frac{\varphi}{2}} S_{\alpha} S_A +\d \zb \ e^{- \frac{\tilde{\varphi}}{2}} \tilde{S}_{\alpha} \tilde{S}_A  \right) \right|_{z=-\zb} \\ 
\nn &= \frac{1}{2 \pi i} \int \left( i \d y \ e^{- \frac{\varphi}{2}} S_{\alpha} S_A + \epsilon' (-i) \d y \ e^{- \frac{\varphi}{2}} S_{\alpha} S_A \right) \\
& =0 \quad \textrm{for } \epsilon'=+1 \ .
\end{align}
Consistency thus implies $\epsilon'=+1$.\footnote{If we would have chosen $\epsilon=+1$ instead, the relevant charge would be $Q_{\alpha A} - \tilde{Q}_{\alpha A}$ and we would require $\epsilon'=-1$ for $Q_{\alpha A} - \tilde{Q}_{\alpha A}$ to be preserved by the D3.}

Having determined $\epsilon'$ we can now derive the behaviour of $V_{\theta}^{-\frac{1}{2}}(z)$ under shifts $z \rightarrow z+1$. Writing $z=x+iy$ and $z' = \frac{1}{2} + x' + iy$ and using the conditions \eqref{eq:Dm1susycond} and \eqref{eq:D3susycond} we find
\begin{align}
\label{eq:VBCDm1} V_{\theta}^{-\frac{1}{2}}(z') &=  e^{- \frac{\varphi}{2}} \ S_{\alpha} S_A ({1}/{2} + x' + i y) = - e^{- \frac{\varphi}{2}} \ \tilde{S}_{\alpha} \tilde{S}_A({1}/{2} - x' + i y) \ , \\
\label{eq:VBCD3} V_{\theta}^{-\frac{1}{2}}(z) &= e^{- \frac{\varphi}{2}} \ S_{\alpha} S_A (x + i y) \qquad  \quad = + e^{- \frac{\varphi}{2}} \ \tilde{S}_{\alpha} \tilde{S}_A(-x + i y) \ .
\end{align}
In the last step we set $x'=x+\frac{1}{2}$. Comparing \eqref{eq:VBCDm1} and \eqref{eq:VBCD3} we conclude that
\be
V_{\theta}^{-\frac{1}{2}}(z+1) = - V_{\theta}^{-\frac{1}{2}}(z) \ .
\ee
The amplitude thus has to be antiperiodic under a shift $z \rightarrow z +1$.
\end{itemize}
We can summarise our findings as follows:
\begin{align}
\label{eq:trafocond1} \mathcal{A}_{D3-E(-1)} & \underset{z_4 \rightarrow z_4+1}{\longrightarrow} - \mathcal{A}_{D3-E(-1)} \ , \\
\label{eq:trafocond2} \mathcal{A}_{D3-E(-1)} & \underset{z_4 \rightarrow z_4+\tau}{\longrightarrow} + \mathcal{A}_{D3-E(-1)} \ .
\end{align}
While the above behaviour applies to the whole amplitude, the only dependence on vertex operator positions $z_i$ enters $\mathcal{A}_{D3-E(-1)}$ through the fermionic and ghost correlator. Thus the behaviour of $\mathcal{A}_{D3-E(-1)}$ under shifts of $z_i$ is completely determined by the transformation properties of \eqref{eq:D3Dm1prelim2}.

\subsection*{Determining the phases $\delta_{\alpha \beta}$ in the sum over spin structures}
We begin by analysing the behaviour of \eqref{eq:D3Dm1prelim2} under $z_4 \rightarrow z_4 +1$. The spin-structure independent part behaves as 
\begin{equation}
{\left( \frac{\vartheta_1(z_{25})}{\vartheta_1^{\prime}(0)} \right)}^{-1} {\left( \frac{\vartheta_1(z_{34})}{\vartheta_1^{\prime}(0)} \right)}^{-1} \underset{z_4 \rightarrow z_4+1}{\longrightarrow} (-1) \ {\left( \frac{\vartheta_1(z_{25})}{\vartheta_1^{\prime}(0)} \right)}^{-1} {\left( \frac{\vartheta_1(z_{34})}{\vartheta_1^{\prime}(0)} \right)}^{-1} \ ,
\end{equation}
where we used \eqref{minus1}. We can use identities \eqref{plushalf} to analyse the transformation of the spin-structure-dependent part of \eqref{eq:D3Dm1prelim2}:
\begin{align}
\nn & \delta_{00} \ \vartheta_{10} \vartheta_{10} \vartheta_{00} \vartheta_{00} + \delta_{01} \ \vartheta_{11} \vartheta_{11} \vartheta_{01} \vartheta_{01} + \delta_{10} \ \vartheta_{00} \vartheta_{00} \vartheta_{10} \vartheta_{10} + \delta_{11} \ \vartheta_{01} \vartheta_{01} \vartheta_{11} \vartheta_{11} \\
\underset{z_4 \rightarrow z_4+1}{\longrightarrow} &  \delta_{00} \ \vartheta_{11} \vartheta_{11} \vartheta_{01} \vartheta_{01} + \delta_{01} \ \vartheta_{10} \vartheta_{10} \vartheta_{00} \vartheta_{00} + \delta_{10} \ \vartheta_{01} \vartheta_{01} \vartheta_{11} \vartheta_{11} + \delta_{11} \ \vartheta_{00} \vartheta_{00} \vartheta_{10} \vartheta_{10} \ ,
\end{align}
where we suppressed the arguments of the $\vartheta$-functions. All in all, the condition \eqref{eq:trafocond1} requires that
\begin{equation}
\label{eq:deltas1} \delta_{00} = + \delta_{01} \ , \qquad \delta_{10} = + \delta_{11} \ .
\end{equation}

Next, we examine transformations under the shift $z_4 \rightarrow z_4 + \tau$. Using \eqref{minustau} we have
\begin{equation}
{\left( \frac{\vartheta_1(z_{25})}{\vartheta_1^{\prime}(0)} \right)}^{-1} {\left( \frac{\vartheta_1(z_{34})}{\vartheta_1^{\prime}(0)} \right)}^{-1} \underset{z_4 \rightarrow z_4+1}{\longrightarrow} (-1) \ e^{\pi i \tau} e^{-2 \pi i z_{34}} \ {\left( \frac{\vartheta_1(z_{25})}{\vartheta_1^{\prime}(0)} \right)}^{-1} {\left( \frac{\vartheta_1(z_{34})}{\vartheta_1^{\prime}(0)} \right)}^{-1} \ .
\end{equation}
We employ \eqref{plushalftau} to transform the spin-structure dependent part:
\begin{align}
\nn & \delta_{00} \ \vartheta_{10} \vartheta_{10} \vartheta_{00} \vartheta_{00} + \delta_{01} \ \vartheta_{11} \vartheta_{11} \vartheta_{01} \vartheta_{01} + \delta_{10} \ \vartheta_{00} \vartheta_{00} \vartheta_{10} \vartheta_{10} + \delta_{11} \ \vartheta_{01} \vartheta_{01} \vartheta_{11} \vartheta_{11} \\
\underset{z_4 \rightarrow z_4+\tau}{\longrightarrow} & e^{- \pi i \tau} e^{2 \pi i z_{34}}  \\ 
\nn \times & \left( \delta_{00} \ \vartheta_{00} \vartheta_{00} \vartheta_{10} \vartheta_{10} + \delta_{01} \ \vartheta_{01} \vartheta_{01} \vartheta_{11} \vartheta_{11} + \delta_{10} \ \vartheta_{10} \vartheta_{10} \vartheta_{00} \vartheta_{00} + \delta_{11} \ \vartheta_{11} \vartheta_{11} \vartheta_{01} \vartheta_{01} \right)\ ,
\end{align}
where we again suppressed the arguments. Putting everything together we find that the factors involving $\tau$ and $z_{34}$ cancel. Thus \eqref{eq:trafocond2} implies that
\begin{equation}
\label{eq:deltas2} \delta_{00} = - \delta_{10} \ , \qquad \delta_{01} = - \delta_{11} \ .
\end{equation}
Then, once we choose a value for one phase, say $\delta_{00}=1$, the two conditions \eqref{eq:deltas1} and \eqref{eq:deltas2} completely fix the remaining phases $\delta_{01}$, $\delta_{10}$ and $\delta_{11}$: 
\begin{equation}
\label{eq:deltas3} \delta_{00} = 1 \ , \qquad \delta_{01} = 1 \ , \qquad \delta_{10} = -1 \ , \qquad \delta_{11}=-1 \ .
\end{equation}
This is conveniently summarised as $\delta_{\alpha \beta} = {(-1)}^{\alpha}$, which is the expression used in \eqref{eq:D3Dm1prelim}. The above analysis can be repeated for the D3-E3 amplitude, where one also finds $\delta_{\alpha \beta} = {(-1)}^{\alpha}$.

\section{Partition function for $\mathcal{N}=2$ sectors}
\label{masspartfuncN2}
Here we evaluate the partition function
\begin{equation}
Z(D3,E3) = \int_{0}^{\infty} \frac{\textrm{d} t}{t} \ \mathcal{Z}_{cl}(r_c,t) = \int_{0}^{\infty} \frac{\d t}{t} \sum_{m,n=-\infty}^{\infty} e^{-\frac{\pi t}{\alpha^{\prime}} \frac{T_2}{U_2} {|m + \bar{U}n +r_c|}^2} \ .
\end{equation}
Integrals of this type have been analysed before in \cite{0404087, 0612110}. 
As a first step it will be useful to Poisson resum the above over $m$ and $n$ to obtain
\begin{equation}
Z(D3,E3) = \frac{\ap}{T_2} \int_{0}^{\infty} \d \ell \sum_{p,q=-\infty}^{\infty} e^{-\frac{\pi \ap}{T_2 U_2} \ell {|q + U p|}^2 + \frac{2 \pi i}{U_2} \textrm{Im} \left(\bar{r}_c [q+Up] \right)} \ ,
\end{equation}
where $\ell =1/t$. Next, we perform the integral over $\ell$. The term $p=q=0$ exhibits a (UV) divergence for $\ell \rightarrow \infty$, which signifies uncancelled tadpoles. In a consistent brane configuration the divergence for $\ell \rightarrow \infty$ will vanish when considering strings stretching between all branes. Hence we exclude the term from the sum. We then arrive at
\begin{equation}
\label{Zpa}
\tilde{Z}(D3,E3) = \frac{U_2}{\pi} \sum_{p,q\neq 0,0} \frac{e^{\frac{2 \pi i}{U_2} \textrm{Im} \left(\bar{r}_c [q+Up] \right)}}{{|q + U p|}^2} \ ,
\end{equation}
where $\tilde{Z}$ is the partition function over massive closed strings only, i.e.~with the term $p=q=0$ removed from the sum. The above also appears in Kronecker's second limit formula:
\begin{equation}
{\left( 2 \ \textrm{Im} (z) \right)}^s \sum_{p,q \neq 0,0} e^{2 \pi i (qu + pv)} \ {|q + p z|}^{-2s} = -4 \pi \ln \left| e^{\pi i u^2 z} \frac{\vartheta_1(v-uz , z)}{\eta(z)} \right| + \mathcal{O}(s-1) \ ,
\end{equation}
and thus the expression \eqref{Zpa} can be written as
\begin{equation}
\tilde{Z}(D3,E3) = - \ln {\left| \frac{\vartheta_1(r_c, U)}{\eta(U)} \right|}^2 + \frac{2 \pi [\textrm{Im}(r_c)]^2}{U_2} = \mathcal{G}(r_c,\bar{r}_c;0,0) \ ,
\end{equation}
where we identified the result as the Greens function on the torus $\mathcal{G}(w_1, \bar{w}_1;w_2, \bar{w}_2)$.

\bibliography{E3deseqbib}  

\providecommand{\href}[2]{#2}\begingroup\raggedright\begin{thebibliography}{10}

\bibitem{9604030}
E.~Witten, {\it {Nonperturbative superpotentials in string theory}},  {\em
  Nucl.Phys.} {\bf B474} (1996) 343--360,
  [\href{http://xxx.lanl.gov/abs/hep-th/9604030}{{\tt hep-th/9604030}}].

\bibitem{KKLT}
S.~Kachru, R.~Kallosh, A.~D. Linde, and S.~P. Trivedi, {\it {De Sitter vacua in
  string theory}},  {\em Phys.Rev.} {\bf D68} (2003) 046005,
  [\href{http://xxx.lanl.gov/abs/hep-th/0301240}{{\tt hep-th/0301240}}].

\bibitem{LVS}
V.~Balasubramanian, P.~Berglund, J.~P. Conlon, and F.~Quevedo, {\it
  {Systematics of moduli stabilisation in Calabi-Yau flux compactifications}},
  {\em JHEP} {\bf 0503} (2005) 007,
  [\href{http://xxx.lanl.gov/abs/hep-th/0502058}{{\tt hep-th/0502058}}].

\bibitem{0609191}
R.~Blumenhagen, M.~Cvetic, and T.~Weigand, {\it {Spacetime instanton
  corrections in 4D string vacua: The Seesaw mechanism for D-Brane models}},
  {\em Nucl.Phys.} {\bf B771} (2007) 113--142,
  [\href{http://xxx.lanl.gov/abs/hep-th/0609191}{{\tt hep-th/0609191}}].

\bibitem{0609213}
L.~Ibanez and A.~Uranga, {\it {Neutrino Majorana Masses from String Theory
  Instanton Effects}},  {\em JHEP} {\bf 0703} (2007) 052,
  [\href{http://xxx.lanl.gov/abs/hep-th/0609213}{{\tt hep-th/0609213}}].

\bibitem{0610003}
B.~Florea, S.~Kachru, J.~McGreevy, and N.~Saulina, {\it {Stringy Instantons and
  Quiver Gauge Theories}},  {\em JHEP} {\bf 0705} (2007) 024,
  [\href{http://xxx.lanl.gov/abs/hep-th/0610003}{{\tt hep-th/0610003}}].

\bibitem{Argurio:2007vqa}
R.~Argurio, M.~Bertolini, G.~Ferretti, A.~Lerda, and C.~Petersson, {\it
  {Stringy instantons at orbifold singularities}},  {\em JHEP} {\bf 06} (2007)
  067, [\href{http://xxx.lanl.gov/abs/0704.0262}{{\tt arXiv:0704.0262}}].

\bibitem{07040784}
M.~Bianchi, F.~Fucito, and J.~F. Morales, {\it {D-brane instantons on the T**6
  / Z(3) orientifold}},  {\em JHEP} {\bf 07} (2007) 038,
  [\href{http://xxx.lanl.gov/abs/0704.0784}{{\tt arXiv:0704.0784}}].

\bibitem{0609211}
M.~Haack, D.~Krefl, D.~Lust, A.~Van~Proeyen, and M.~Zagermann, {\it {Gaugino
  Condensates and D-terms from D7-branes}},  {\em JHEP} {\bf 0701} (2007) 078,
  [\href{http://xxx.lanl.gov/abs/hep-th/0609211}{{\tt hep-th/0609211}}].

\bibitem{09023251}
R.~Blumenhagen, M.~Cvetic, S.~Kachru, and T.~Weigand, {\it {D-Brane Instantons
  in Type II Orientifolds}},  {\em Ann.Rev.Nucl.Part.Sci.} {\bf 59} (2009)
  269--296, [\href{http://xxx.lanl.gov/abs/0902.3251}{{\tt arXiv:0902.3251}}].

\bibitem{0612110}
S.~A. Abel and M.~D. Goodsell, {\it {Realistic Yukawa Couplings through
  Instantons in Intersecting Brane Worlds}},  {\em JHEP} {\bf 0710} (2007) 034,
  [\href{http://xxx.lanl.gov/abs/hep-th/0612110}{{\tt hep-th/0612110}}].

\bibitem{07071871}
R.~Blumenhagen, M.~Cvetic, D.~Lust, R.~Richter, and T.~Weigand, {\it
  {Non-perturbative Yukawa Couplings from String Instantons}},  {\em
  Phys.Rev.Lett.} {\bf 100} (2008) 061602,
  [\href{http://xxx.lanl.gov/abs/0707.1871}{{\tt arXiv:0707.1871}}].

\bibitem{07111316}
L.~Ibanez and A.~Uranga, {\it {Instanton induced open string superpotentials
  and branes at singularities}},  {\em JHEP} {\bf 0802} (2008) 103,
  [\href{http://xxx.lanl.gov/abs/0711.1316}{{\tt arXiv:0711.1316}}].

\bibitem{08111583}
L.~Ibanez and R.~Richter, {\it {Stringy Instantons and Yukawa Couplings in
  MSSM-like Orientifold Models}},  {\em JHEP} {\bf 0903} (2009) 090,
  [\href{http://xxx.lanl.gov/abs/0811.1583}{{\tt arXiv:0811.1583}}].

\bibitem{09053379}
M.~Cvetic, J.~Halverson, and R.~Richter, {\it {Realistic Yukawa structures from
  orientifold compactifications}},  {\em JHEP} {\bf 0912} (2009) 063,
  [\href{http://xxx.lanl.gov/abs/0905.3379}{{\tt arXiv:0905.3379}}].

\bibitem{0703028}
M.~Cvetic, R.~Richter, and T.~Weigand, {\it {Computation of D-brane instanton
  induced superpotential couplings: Majorana masses from string theory}},  {\em
  Phys.Rev.} {\bf D76} (2007) 086002,
  [\href{http://xxx.lanl.gov/abs/hep-th/0703028}{{\tt hep-th/0703028}}].

\bibitem{07041079}
L.~Ibanez, A.~Schellekens, and A.~Uranga, {\it {Instanton Induced Neutrino
  Majorana Masses in CFT Orientifolds with MSSM-like spectra}},  {\em JHEP}
  {\bf 0706} (2007) 011, [\href{http://xxx.lanl.gov/abs/0704.1079}{{\tt
  arXiv:0704.1079}}].

\bibitem{Blumenhagen:2008ji}
R.~Blumenhagen and M.~Schmidt-Sommerfeld, {\it {Power Towers of String
  Instantons for N=1 Vacua}},  {\em JHEP} {\bf 07} (2008) 027,
  [\href{http://xxx.lanl.gov/abs/0803.1562}{{\tt arXiv:0803.1562}}].

\bibitem{Blumenhagen:2012kz}
R.~Blumenhagen, X.~Gao, T.~Rahn, and P.~Shukla, {\it {A Note on Poly-Instanton
  Effects in Type IIB Orientifolds on Calabi-Yau Threefolds}},  {\em JHEP} {\bf
  06} (2012) 162, [\href{http://xxx.lanl.gov/abs/1205.2485}{{\tt
  arXiv:1205.2485}}].

\bibitem{Cicoli:2011ct}
M.~Cicoli, F.~G. Pedro, and G.~Tasinato, {\it {Poly-instanton Inflation}},
  {\em JCAP} {\bf 1112} (2011) 022,
  [\href{http://xxx.lanl.gov/abs/1110.6182}{{\tt arXiv:1110.6182}}].

\bibitem{Cicoli:2012tz}
M.~Cicoli, F.~G. Pedro, and G.~Tasinato, {\it {Natural Quintessence in String
  Theory}},  {\em JCAP} {\bf 1207} (2012) 044,
  [\href{http://xxx.lanl.gov/abs/1203.6655}{{\tt arXiv:1203.6655}}].

\bibitem{12031750}
M.~Cicoli, A.~Maharana, F.~Quevedo, and C.~P. Burgess, {\it {De Sitter String
  Vacua from Dilaton-dependent Non-perturbative Effects}},  {\em JHEP} {\bf 06}
  (2012) 011, [\href{http://xxx.lanl.gov/abs/1203.1750}{{\tt
  arXiv:1203.1750}}].

\bibitem{08062291}
D.~Forcella, I.~Garcia-Etxebarria, and A.~Uranga, {\it {E3-brane instantons and
  baryonic operators for D3-branes on toric singularities}},  {\em JHEP} {\bf
  0903} (2009) 041, [\href{http://xxx.lanl.gov/abs/0806.2291}{{\tt
  arXiv:0806.2291}}].

\bibitem{09105496}
F.~Marchesano and L.~Martucci, {\it {Non-perturbative effects on seven-brane
  Yukawa couplings}},  {\em Phys.Rev.Lett.} {\bf 104} (2010) 231601,
  [\href{http://xxx.lanl.gov/abs/0910.5496}{{\tt arXiv:0910.5496}}].

\bibitem{12116529}
A.~Font, L.~E. Ibanez, F.~Marchesano, and D.~Regalado, {\it {Non-perturbative
  effects and Yukawa hierarchies in F-theory SU(5) Unification}},  {\em JHEP}
  {\bf 1303} (2013) 140, [\href{http://xxx.lanl.gov/abs/1211.6529}{{\tt
  arXiv:1211.6529}}].

\bibitem{13078089}
A.~Font, F.~Marchesano, D.~Regalado, and G.~Zoccarato, {\it {Up-type quark
  masses in SU(5) F-theory models}},  {\em JHEP} {\bf 1311} (2013) 125,
  [\href{http://xxx.lanl.gov/abs/1307.8089}{{\tt arXiv:1307.8089}}].

\bibitem{150302683}
F.~Marchesano, D.~Regalado, and G.~Zoccarato, {\it {Yukawa hierarchies at the
  point of $E_8$ in F-theory}},  \href{http://xxx.lanl.gov/abs/1503.0268}{{\tt
  arXiv:1503.0268}}.

\bibitem{0607050}
D.~Baumann, A.~Dymarsky, I.~R. Klebanov, J.~M. Maldacena, L.~P. McAllister,
  et~al., {\it {On D3-brane Potentials in Compactifications with Fluxes and
  Wrapped D-branes}},  {\em JHEP} {\bf 0611} (2006) 031,
  [\href{http://xxx.lanl.gov/abs/hep-th/0607050}{{\tt hep-th/0607050}}].

\bibitem{0404087}
M.~Berg, M.~Haack, and B.~Kors, {\it {Loop corrections to volume moduli and
  inflation in string theory}},  {\em Phys.Rev.} {\bf D71} (2005) 026005,
  [\href{http://xxx.lanl.gov/abs/hep-th/0404087}{{\tt hep-th/0404087}}].

\bibitem{12071103}
M.~Berg, J.~P. Conlon, D.~Marsh, and L.~T. Witkowski, {\it {Superpotential
  de-sequestering in string models}},  {\em JHEP} {\bf 1302} (2013) 018,
  [\href{http://xxx.lanl.gov/abs/1207.1103}{{\tt arXiv:1207.1103}}].

\bibitem{9707209}
A.~Brignole, L.~E. Ibanez, and C.~Munoz, {\it {Soft supersymmetry breaking
  terms from supergravity and superstring models}},  {\em Adv.Ser.Direct.High
  Energy Phys.} {\bf 21} (2010) 244--268,
  [\href{http://xxx.lanl.gov/abs/hep-ph/9707209}{{\tt hep-ph/9707209}}].

\bibitem{0703105}
S.~Kachru, L.~McAllister, and R.~Sundrum, {\it {Sequestering in String
  Theory}},  {\em JHEP} {\bf 0710} (2007) 013,
  [\href{http://xxx.lanl.gov/abs/hep-th/0703105}{{\tt hep-th/0703105}}].

\bibitem{10121858}
M.~Berg, D.~Marsh, L.~McAllister, and E.~Pajer, {\it {Sequestering in String
  Compactifications}},  {\em JHEP} {\bf 1106} (2011) 134,
  [\href{http://xxx.lanl.gov/abs/1012.1858}{{\tt arXiv:1012.1858}}].

\bibitem{Coughlan:1983ci}
G.~Coughlan, W.~Fischler, E.~W. Kolb, S.~Raby, and G.~G. Ross, {\it
  {Cosmological Problems for the Polonyi Potential}},  {\em Phys.Lett.} {\bf
  B131} (1983) 59.

\bibitem{9308292}
T.~Banks, D.~B. Kaplan, and A.~E. Nelson, {\it {Cosmological implications of
  dynamical supersymmetry breaking}},  {\em Phys.Rev.} {\bf D49} (1994)
  779--787, [\href{http://xxx.lanl.gov/abs/hep-ph/9308292}{{\tt
  hep-ph/9308292}}].

\bibitem{9308325}
B.~de~Carlos, J.~Casas, F.~Quevedo, and E.~Roulet, {\it {Model independent
  properties and cosmological implications of the dilaton and moduli sectors of
  4-d strings}},  {\em Phys.Lett.} {\bf B318} (1993) 447--456,
  [\href{http://xxx.lanl.gov/abs/hep-ph/9308325}{{\tt hep-ph/9308325}}].

\bibitem{09063297}
R.~Blumenhagen, J.~Conlon, S.~Krippendorf, S.~Moster, and F.~Quevedo, {\it
  {SUSY Breaking in Local String/F-Theory Models}},  {\em JHEP} {\bf 0909}
  (2009) 007, [\href{http://xxx.lanl.gov/abs/0906.3297}{{\tt
  arXiv:0906.3297}}].

\bibitem{14091931}
L.~Aparicio, M.~Cicoli, S.~Krippendorf, A.~Maharana, F.~Muia, and F.~Quevedo,
  {\it {Sequestered de Sitter String Scenarios: Soft-terms}},  {\em JHEP} {\bf
  11} (2014) 071, [\href{http://xxx.lanl.gov/abs/1409.1931}{{\tt
  arXiv:1409.1931}}].

\bibitem{Akerblom:2007uc}
N.~Akerblom, R.~Blumenhagen, D.~Lust, and M.~Schmidt-Sommerfeld, {\it
  {Instantons and Holomorphic Couplings in Intersecting D-brane Models}},  {\em
  JHEP} {\bf 08} (2007) 044, [\href{http://xxx.lanl.gov/abs/0705.2366}{{\tt
  arXiv:0705.2366}}].

\bibitem{Petersson:2007sc}
C.~Petersson, {\it {Superpotentials From Stringy Instantons Without
  Orientifolds}},  {\em JHEP} {\bf 05} (2008) 078,
  [\href{http://xxx.lanl.gov/abs/0711.1837}{{\tt arXiv:0711.1837}}].

\bibitem{07080403}
R.~Blumenhagen, M.~Cvetic, R.~Richter, and T.~Weigand, {\it {Lifting
  D-Instanton Zero Modes by Recombination and Background Fluxes}},  {\em JHEP}
  {\bf 0710} (2007) 098, [\href{http://xxx.lanl.gov/abs/0708.0403}{{\tt
  arXiv:0708.0403}}].

\bibitem{Billo:2002hm}
M.~Billo, M.~Frau, I.~Pesando, F.~Fucito, A.~Lerda, and A.~Liccardo, {\it
  {Classical gauge instantons from open strings}},  {\em JHEP} {\bf 02} (2003)
  045, [\href{http://xxx.lanl.gov/abs/hep-th/0211250}{{\tt hep-th/0211250}}].

\bibitem{Dine:1986zy}
M.~Dine, N.~Seiberg, X.~G. Wen, and E.~Witten, {\it {Nonperturbative Effects on
  the String World Sheet}},  {\em Nucl. Phys.} {\bf B278} (1986) 769.

\bibitem{Dine:1987bq}
M.~Dine, N.~Seiberg, X.~G. Wen, and E.~Witten, {\it {Nonperturbative Effects on
  the String World Sheet. 2.}},  {\em Nucl. Phys.} {\bf B289} (1987) 319.

\bibitem{Akerblom:2006hx}
N.~Akerblom, R.~Blumenhagen, D.~Lust, E.~Plauschinn, and M.~Schmidt-Sommerfeld,
  {\it {Non-perturbative SQCD Superpotentials from String Instantons}},  {\em
  JHEP} {\bf 04} (2007) 076,
  [\href{http://xxx.lanl.gov/abs/hep-th/0612132}{{\tt hep-th/0612132}}].

\bibitem{Conlon:2009kt}
J.~P. Conlon and E.~Palti, {\it {Gauge Threshold Corrections for Local
  Orientifolds}},  {\em JHEP} {\bf 09} (2009) 019,
  [\href{http://xxx.lanl.gov/abs/0906.1920}{{\tt arXiv:0906.1920}}].

\bibitem{PolVol1}
J.~Polchinski, {\em String Theory: Volume 1, An Introduction to the Bosonic
  String}.
\newblock Cambridge Monographs on Mathematical Physics. Cambridge University
  Press, 1998.

\bibitem{PolVol2}
J.~Polchinski, {\em String Theory: Volume 2, Superstring Theory and Beyond}.
\newblock Cambridge Monographs on Mathematical Physics. Cambridge University
  Press, 2005.

\bibitem{10084361}
J.~P. Conlon, M.~Goodsell, and E.~Palti, {\it {Anomaly Mediation in Superstring
  Theory}},  {\em Fortsch.Phys.} {\bf 59} (2011) 5--75,
  [\href{http://xxx.lanl.gov/abs/1008.4361}{{\tt arXiv:1008.4361}}].

\bibitem{0609180}
J.~P. Conlon, D.~Cremades, and F.~Quevedo, {\it {Kahler potentials of chiral
  matter fields for Calabi-Yau string compactifications}},  {\em JHEP} {\bf
  0701} (2007) 022, [\href{http://xxx.lanl.gov/abs/hep-th/0609180}{{\tt
  hep-th/0609180}}].

\bibitem{9603167}
M.~R. Douglas and G.~W. Moore, {\it {D-branes, quivers, and ALE instantons}},
  \href{http://xxx.lanl.gov/abs/hep-th/9603167}{{\tt hep-th/9603167}}.

\bibitem{9704151}
M.~R. Douglas, B.~R. Greene, and D.~R. Morrison, {\it {Orbifold resolution by
  D-branes}},  {\em Nucl.Phys.} {\bf B506} (1997) 84--106,
  [\href{http://xxx.lanl.gov/abs/hep-th/9704151}{{\tt hep-th/9704151}}].

\bibitem{11106454}
J.~P. Conlon, {\it {Brane-Antibrane Backreaction in Axion Monodromy
  Inflation}},  {\em JCAP} {\bf 1201} (2012) 033,
  [\href{http://xxx.lanl.gov/abs/1110.6454}{{\tt arXiv:1110.6454}}].

\bibitem{07103883}
T.~W. Grimm, {\it {Axion inflation in type II string theory}},  {\em Phys.Rev.}
  {\bf D77} (2008) 126007, [\href{http://xxx.lanl.gov/abs/0710.3883}{{\tt
  arXiv:0710.3883}}].

\bibitem{Mumford1}
D.~Mumford, C.~Musili, E.~Previato, and M.~Stillmann, {\em Tata Lectures on
  Theta I}.
\newblock Modern Birkh{\"a}user Classics. Springer, London, 2007.

\bibitem{0404134}
D.~Lust, P.~Mayr, R.~Richter, and S.~Stieberger, {\it {Scattering of gauge,
  matter, and moduli fields from intersecting branes}},  {\em Nucl.Phys.} {\bf
  B696} (2004) 205--250, [\href{http://xxx.lanl.gov/abs/hep-th/0404134}{{\tt
  hep-th/0404134}}].

\bibitem{0211250}
M.~Billo, M.~Frau, I.~Pesando, F.~Fucito, A.~Lerda, et~al., {\it {Classical
  gauge instantons from open strings}},  {\em JHEP} {\bf 0302} (2003) 045,
  [\href{http://xxx.lanl.gov/abs/hep-th/0211250}{{\tt hep-th/0211250}}].

\end{thebibliography}\endgroup
\bibliographystyle{JHEP}
\end{document}